
\catcode`\@=11
\font\tensmc=cmcsc10      
\def\smc{\tensmc}

\def\hcorrection#1{\advance\hoffset by #1 }
\def\vcorrection#1{\advance\voffset by #1 }
\def\wlog#1{}
\newif\iftitle@
\outer\def\title{\title@true\vglue 24\p@ plus 12\p@ minus 12\p@
   \bgroup\let\\=\cr\tabskip\centering
   \halign to \hsize\bgroup\tenbf\hfill\ignorespaces##\unskip\hfill\cr}
\def\endtitle{\cr\egroup\egroup\vglue 18\p@ plus 12\p@ minus 6\p@}
\outer\def\author{\iftitle@\vglue -18\p@ plus -12\p@ minus -6\p@\fi\vglue
    12\p@ plus 6\p@ minus 3\p@\bgroup\let\\=\cr\tabskip\centering
    \halign to \hsize\bgroup\smc\hfill\ignorespaces##\unskip\hfill\cr}
\def\endauthor{\cr\egroup\egroup\vglue 18\p@ plus 12\p@ minus 6\p@}
\outer\def\heading{\bigbreak\bgroup\let\\=\cr\tabskip\centering
    \halign to \hsize\bgroup\smc\hfill\ignorespaces##\unskip\hfill\cr}
\def\endheading{\cr\egroup\egroup\nobreak\medskip}

\outer\def\endproclaim{\par\ifdim\lastskip<\medskipamount\removelastskip
  \penalty 55 \fi\medskip\rm}
\outer\def\demo#1{\par\ifdim\lastskip<\smallskipamount\removelastskip
    \smallskip\fi\noindent{\smc\ignorespaces#1\unskip:\enspace}\rm
      \ignorespaces}

\newcount\footmarkcount@
\footmarkcount@=1
\def\makefootnote@#1#2{\insert\footins{\interlinepenalty=100
  \splittopskip=\ht\strutbox \splitmaxdepth=\dp\strutbox
  \floatingpenalty=\@MM
  \leftskip=\z@\rightskip=\z@\spaceskip=\z@\xspaceskip=\z@
  \noindent{#1}\footstrut\rm\ignorespaces #2\strut}}
\def\footnote{\let\@sf=\empty\ifhmode\edef\@sf{\spacefactor
   =\the\spacefactor}\/\fi\futurelet\next\footnote@}
\def\footnote@{\ifx"\next\let\next\footnote@@\else
    \let\next\footnote@@@\fi\next}
\def\footnote@@"#1"#2{#1\@sf\relax\makefootnote@{#1}{#2}}
\def\footnote@@@#1{$^{\number\footmarkcount@}$\makefootnote@
   {$^{\number\footmarkcount@}$}{#1}\global\advance\footmarkcount@ by 1 }

\hyphenation{man-u-script man-u-scripts ap-pen-dix ap-pen-di-ces}
\hyphenation{data-base data-bases}
\ifx\amstexloaded@\relax\catcode`\@=13
  \endinput\else\let\amstexloaded@=\relax\fi
\newlinechar=`\^^J
\def\eat@#1{}
\def\Space@.{\futurelet\Space@\relax}
\Space@. %
\newhelp\athelp@
{Only certain combinations beginning with @ make sense to me.^^J
Perhaps you wanted \string\@\space for a printed @?^^J
I've ignored the character or group after @.}
\def\futureletnextat@{\futurelet\next\at@}
{\catcode`\@=\active
\lccode`\Z=`\@ \lowercase
{\gdef@{\expandafter\csname futureletnextatZ\endcsname}
\expandafter\gdef\csname atZ\endcsname
   {\ifcat\noexpand\next a\def\next{\csname atZZ\endcsname}\else
   \ifcat\noexpand\next0\def\next{\csname atZZ\endcsname}\else
    \def\next{\csname atZZZ\endcsname}\fi\fi\next}
\expandafter\gdef\csname atZZ\endcsname#1{\expandafter
   \ifx\csname #1Zat\endcsname\relax\def\next
     {\errhelp\expandafter=\csname athelpZ\endcsname
      \errmessage{Invalid use of \string@}}\else
       \def\next{\csname #1Zat\endcsname}\fi\next}
\expandafter\gdef\csname atZZZ\endcsname#1{\errhelp
    \expandafter=\csname athelpZ\endcsname
      \errmessage{Invalid use of \string@}}}}
\def\atdef@#1{\expandafter\def\csname #1@at\endcsname}
\newhelp\defahelp@{If you typed \string\define\space cs instead of
\string\define\string\cs\space^^J
I've substituted an inaccessible control sequence so that your^^J
definition will be completed without mixing me up too badly.^^J
If you typed \string\define{\string\cs} the inaccessible control sequence^^J
was defined to be \string\cs, and the rest of your^^J
definition appears as input.}
\newhelp\defbhelp@{I've ignored your definition, because it might^^J
conflict with other uses that are important to me.}
\def\define{\futurelet\next\define@}
\def\define@{\ifcat\noexpand\next\relax
  \def\next{\define@@}%
  \else\errhelp=\defahelp@
  \errmessage{\string\define\space must be followed by a control
     sequence}\def\next{\def\garbage@}\fi\next}
\def\undefined@{}
\def\preloaded@{}
\def\define@@#1{\ifx#1\relax\errhelp=\defbhelp@
   \errmessage{\string#1\space is already defined}\def\next{\def\garbage@}%
   \else\expandafter\ifx\csname\expandafter\eat@\string
         #1@\endcsname\undefined@\errhelp=\defbhelp@
   \errmessage{\string#1\space can't be defined}\def\next{\def\garbage@}%
   \else\expandafter\ifx\csname\expandafter\eat@\string#1\endcsname\relax
     \def\next{\def#1}\else\errhelp=\defbhelp@
     \errmessage{\string#1\space is already defined}\def\next{\def\garbage@}%
      \fi\fi\fi\next}
\def\famzero{\fam\z@}

\def\det{\mathop{\famzero det}}

\def\exp{\mathop{\famzero exp}\nolimits}

\def\lim{\mathop{\famzero lim}}

\def\ln{\mathop{\famzero ln}\nolimits}

\def\textfont@#1#2{\def#1{\relax\ifmmode
    \errmessage{Use \string#1\space only in text}\else#2\fi}}
\textfont@\rm\tenrm
\textfont@\it\tenit
\textfont@\sl\tensl
\textfont@\bf\tenbf
\textfont@\smc\tensmc
\let\ic@=\/
\def\/{\unskip\ic@}
\def\textfonti{\the\textfont1 }
\def\t#1#2{{\edef\next{\the\font}\textfonti\accent"7F \next#1#2}}
\let\B=\=
\let\D=\.
\def~{\unskip\nobreak\ \ignorespaces}
{\catcode`\@=\active
\gdef\@{\char'100 }}
\atdef@-{\leavevmode\futurelet\next\athyph@}
\def\athyph@{\ifx\next-\let\next=\athyph@@
  \else\let\next=\athyph@@@\fi\next}
\def\athyph@@@{\hbox{-}}
\def\athyph@@#1{\futurelet\next\athyph@@@@}
\def\athyph@@@@{\if\next-\def\next##1{\hbox{---}}\else
    \def\next{\hbox{--}}\fi\next}
\def\.{.\spacefactor=\@m}
\atdef@.{\null.}
\atdef@,{\null,}
\atdef@;{\null;}
\atdef@:{\null:}
\atdef@?{\null?}
\atdef@!{\null!}
\def\srdr@{\thinspace}
\def\drsr@{\kern.02778em}
\def\sldl@{\kern.02778em}
\def\dlsl@{\thinspace}
\atdef@"{\unskip\futurelet\next\atqq@}
\def\atqq@{\ifx\next\Space@\def\next. {\atqq@@}\else
         \def\next.{\atqq@@}\fi\next.}
\def\atqq@@{\futurelet\next\atqq@@@}
\def\atqq@@@{\ifx\next`\def\next`{\atqql@}\else\def\next'{\atqqr@}\fi\next}
\def\atqql@{\futurelet\next\atqql@@}
\def\atqql@@{\ifx\next`\def\next`{\sldl@``}\else\def\next{\dlsl@`}\fi\next}
\def\atqqr@{\futurelet\next\atqqr@@}
\def\atqqr@@{\ifx\next'\def\next'{\srdr@''}\else\def\next{\drsr@'}\fi\next}
\def\flushpar{\par\noindent}
\def\textfontii{\the\textfont2 }
\def\{{\relax\ifmmode\lbrace\else
    {\textfontii f}\spacefactor=\@m\fi}
\def\}{\relax\ifmmode\rbrace\else
    \let\@sf=\empty\ifhmode\edef\@sf{\spacefactor=\the\spacefactor}\fi
      {\textfontii g}\@sf\relax\fi}
\def\nonhmodeerr@#1{\errmessage
     {\string#1\space allowed only within text}}
\def\linebreak{\relax\ifhmode\unskip\break\else
    \nonhmodeerr@\linebreak\fi}
\def\allowlinebreak{\relax
   \ifhmode\allowbreak\else\nonhmodeerr@\allowlinebreak\fi}
\newskip\saveskip@
\def\nolinebreak{\relax\ifhmode\saveskip@=\lastskip\unskip
  \nobreak\ifdim\saveskip@>\z@\hskip\saveskip@\fi
   \else\nonhmodeerr@\nolinebreak\fi}
\def\newline{\relax\ifhmode\null\hfil\break
    \else\nonhmodeerr@\newline\fi}
\def\nonmathaerr@#1{\errmessage
     {\string#1\space is not allowed in display math mode}}
\def\nonmathberr@#1{\errmessage{\string#1\space is allowed only in math mode}}
\def\mathbreak{\relax\ifmmode\ifinner\break\else
   \nonmathaerr@\mathbreak\fi\else\nonmathberr@\mathbreak\fi}
\def\nomathbreak{\relax\ifmmode\ifinner\nobreak\else
    \nonmathaerr@\nomathbreak\fi\else\nonmathberr@\nomathbreak\fi}
\def\allowmathbreak{\relax\ifmmode\ifinner\allowbreak\else
     \nonmathaerr@\allowmathbreak\fi\else\nonmathberr@\allowmathbreak\fi}
\def\pagebreak{\relax\ifmmode
   \ifinner\errmessage{\string\pagebreak\space
     not allowed in non-display math mode}\else\postdisplaypenalty-\@M\fi
   \else\ifvmode\penalty-\@M\else\edef\spacefactor@
       {\spacefactor=\the\spacefactor}\vadjust{\penalty-\@M}\spacefactor@
        \relax\fi\fi}
\def\nopagebreak{\relax\ifmmode
     \ifinner\errmessage{\string\nopagebreak\space
    not allowed in non-display math mode}\else\postdisplaypenalty\@M\fi
    \else\ifvmode\nobreak\else\edef\spacefactor@
        {\spacefactor=\the\spacefactor}\vadjust{\penalty\@M}\spacefactor@
         \relax\fi\fi}
\def\newpage{\relax\ifvmode\vfill\penalty-\@M\else\nonvmodeerr@\newpage\fi}
\def\nonvmodeerr@#1{\errmessage
    {\string#1\space is allowed only between paragraphs}}
\def\smallpagebreak{\relax\ifvmode\smallbreak
      \else\nonvmodeerr@\smallpagebreak\fi}
\def\medpagebreak{\relax\ifvmode\medbreak
       \else\nonvmodeerr@\medpagebreak\fi}
\def\bigpagebreak{\relax\ifvmode\bigbreak
      \else\nonvmodeerr@\bigpagebreak\fi}
\newdimen\captionwidth@
\captionwidth@=\hsize
\advance\captionwidth@ by -1.5in
\def\caption#1{}
\def\topspace#1{\gdef\thespace@{#1}\ifvmode\def\next
    {\futurelet\next\topspace@}\else\def\next{\nonvmodeerr@\topspace}\fi\next}
\def\topspace@{\ifx\next\Space@\def\next. {\futurelet\next\topspace@@}\else
     \def\next.{\futurelet\next\topspace@@}\fi\next.}
\def\topspace@@{\ifx\next\caption\let\next\topspace@@@\else
    \let\next\topspace@@@@\fi\next}
 \def\topspace@@@@{\topinsert\vbox to
       \thespace@{}\endinsert}
\def\topspace@@@\caption#1{\topinsert\vbox to
    \thespace@{}\nobreak
      \smallskip
    \setbox\z@=\hbox{\noindent\ignorespaces#1\unskip}%
   \ifdim\wd\z@>\captionwidth@
   \centerline{\vbox{\hsize=\captionwidth@\noindent\ignorespaces#1\unskip}}%
   \else\centerline{\box\z@}\fi\endinsert}
\def\midspace#1{\gdef\thespace@{#1}\ifvmode\def\next
    {\futurelet\next\midspace@}\else\def\next{\nonvmodeerr@\midspace}\fi\next}
\def\midspace@{\ifx\next\Space@\def\next. {\futurelet\next\midspace@@}\else
     \def\next.{\futurelet\next\midspace@@}\fi\next.}
\def\midspace@@{\ifx\next\caption\let\next\midspace@@@\else
    \let\next\midspace@@@@\fi\next}
 \def\midspace@@@@{\midinsert\vbox to
       \thespace@{}\endinsert}
\def\midspace@@@\caption#1{\midinsert\vbox to
    \thespace@{}\nobreak
      \smallskip
      \setbox\z@=\hbox{\noindent\ignorespaces#1\unskip}%
      \ifdim\wd\z@>\captionwidth@
    \centerline{\vbox{\hsize=\captionwidth@\noindent\ignorespaces#1\unskip}}%
    \else\centerline{\box\z@}\fi\endinsert}
\mathchardef\prime@="0230
\def\prime{{{}\prime@{}}}
\def\prim@s{\prime@\futurelet\next\pr@m@s}
\let\dsize=\displaystyle

\def\,{\relax\ifmmode\mskip\thinmuskip\else\thinspace\fi}
\def\!{\relax\ifmmode\mskip-\thinmuskip\else\negthinspace\fi}
\def\frac#1#2{{#1\over#2}}
\def\dfrac#1#2{{\displaystyle{#1\over#2}}}
\def\tfrac#1#2{{\textstyle{#1\over#2}}}

\def\:{\nobreak\hskip.1111em{:}\hskip.3333em plus .0555em\relax}
\def\intic@{\mathchoice{\hskip5\p@}{\hskip4\p@}{\hskip4\p@}{\hskip4\p@}}
\def\negintic@
 {\mathchoice{\hskip-5\p@}{\hskip-4\p@}{\hskip-4\p@}{\hskip-4\p@}}
\def\intkern@{\mathchoice{\!\!\!}{\!\!}{\!\!}{\!\!}}
\def\intdots@{\mathchoice{\cdots}{{\cdotp}\mkern1.5mu
    {\cdotp}\mkern1.5mu{\cdotp}}{{\cdotp}\mkern1mu{\cdotp}\mkern1mu
      {\cdotp}}{{\cdotp}\mkern1mu{\cdotp}\mkern1mu{\cdotp}}}
\newcount\intno@
\def\iint{\intno@=\tw@\futurelet\next\ints@}
\def\iiint{\intno@=\thr@@\futurelet\next\ints@}
\def\iiiint{\intno@=4 \futurelet\next\ints@}
\def\idotsint{\intno@=\z@\futurelet\next\ints@}
\def\ints@{\findlimits@\ints@@}
\newif\iflimtoken@
\newif\iflimits@
\def\findlimits@{\limtoken@false\limits@false\ifx\next\limits
 \limtoken@true\limits@true\else\ifx\next\nolimits\limtoken@true\limits@false
    \fi\fi}
\def\multintlimits@{\intop\ifnum\intno@=\z@\intdots@
  \else\intkern@\fi
    \ifnum\intno@>\tw@\intop\intkern@\fi
     \ifnum\intno@>\thr@@\intop\intkern@\fi\intop}
\def\multint@{\int\ifnum\intno@=\z@\intdots@\else\intkern@\fi
   \ifnum\intno@>\tw@\int\intkern@\fi
    \ifnum\intno@>\thr@@\int\intkern@\fi\int}
\def\ints@@{\iflimtoken@\def\ints@@@{\iflimits@
   \negintic@\mathop{\intic@\multintlimits@}\limits\else
    \multint@\nolimits\fi\eat@}\else
     \def\ints@@@{\multint@\nolimits}\fi\ints@@@}
\def\Sb{_\bgroup\vspace@
        \baselineskip=\fontdimen10 \scriptfont\tw@
        \advance\baselineskip by \fontdimen12 \scriptfont\tw@
        \lineskip=\thr@@\fontdimen8 \scriptfont\thr@@
        \lineskiplimit=\thr@@\fontdimen8 \scriptfont\thr@@
        \Let@\vbox\bgroup\halign\bgroup \hfil$\scriptstyle
            {##}$\hfil\cr}
\def\endSb{\crcr\egroup\egroup\egroup}
\def\Sp{^\bgroup\vspace@
        \baselineskip=\fontdimen10 \scriptfont\tw@
        \advance\baselineskip by \fontdimen12 \scriptfont\tw@
        \lineskip=\thr@@\fontdimen8 \scriptfont\thr@@
        \lineskiplimit=\thr@@\fontdimen8 \scriptfont\thr@@
        \Let@\vbox\bgroup\halign\bgroup \hfil$\scriptstyle
            {##}$\hfil\cr}
\def\endSp{\crcr\egroup\egroup\egroup}
\def\Let@{\relax\iffalse{\fi\let\\=\cr\iffalse}\fi}
\def\vspace@{\def\vspace##1{\noalign{\vskip##1 }}}
\def\aligned{\,\vcenter\bgroup\vspace@\Let@\openup\jot\m@th\ialign
  \bgroup \strut\hfil$\displaystyle{##}$&$\displaystyle{{}##}$\hfil\crcr}
\def\endaligned{\crcr\egroup\egroup}
\def\matrix{\,\vcenter\bgroup\Let@\vspace@
    \normalbaselines
  \m@th\ialign\bgroup\hfil$##$\hfil&&\quad\hfil$##$\hfil\crcr
    \mathstrut\crcr\noalign{\kern-\baselineskip}}
\def\endmatrix{\crcr\mathstrut\crcr\noalign{\kern-\baselineskip}\egroup
                \egroup\,}
\newtoks\hashtoks@
\hashtoks@={#}
\def\format{\crcr\egroup\iffalse{\fi\ifnum`}=0 \fi\format@}
\def\format@#1\\{\def\preamble@{#1}%
  \def\c{\hfil$\the\hashtoks@$\hfil}%
  \def\r{\hfil$\the\hashtoks@$}%
  \def\l{$\the\hashtoks@$\hfil}%
  \setbox\z@=\hbox{\xdef\Preamble@{\preamble@}}\ifnum`{=0 \fi\iffalse}\fi
   \ialign\bgroup\span\Preamble@\crcr}

\def\cases{\left\{\,\vcenter\bgroup\vspace@
     \normalbaselines\openup\jot\m@th
       \Let@\ialign\bgroup$##$\hfil&\quad$##$\hfil\crcr
      \mathstrut\crcr\noalign{\kern-\baselineskip}}

\newif\iftagsleft@
\tagsleft@true
\def\TagsOnRight{\global\tagsleft@false}
\def\tag#1$${\iftagsleft@\leqno\else\eqno\fi
 \hbox{\def\pagebreak{\global\postdisplaypenalty-\@M}%
 \def\nopagebreak{\global\postdisplaypenalty\@M}\rm(#1\unskip)}%
  $$\postdisplaypenalty\z@\ignorespaces}
\interdisplaylinepenalty=\@M
\def\allowdisplaybreak@{\def\allowdisplaybreak{\noalign{\allowbreak}}}
\def\displaybreak@{\def\displaybreak{\noalign{\break}}}
\def\align#1\endalign{\def\tag{&}\vspace@\allowdisplaybreak@\displaybreak@
  \iftagsleft@\lalign@#1\endalign\else
   \ralign@#1\endalign\fi}
\def\ralign@#1\endalign{\displ@y\Let@\tabskip\centering\halign to\displaywidth
     {\hfil$\displaystyle{##}$\tabskip=\z@&$\displaystyle{{}##}$\hfil
       \tabskip=\centering&\llap{\hbox{(\rm##\unskip)}}\tabskip\z@\crcr
             #1\crcr}}
\def\lalign@
 #1\endalign{\displ@y\Let@\tabskip\centering\halign to \displaywidth
   {\hfil$\displaystyle{##}$\tabskip=\z@&$\displaystyle{{}##}$\hfil
   \tabskip=\centering&\kern-\displaywidth
        \rlap{\hbox{(\rm##\unskip)}}\tabskip=\displaywidth\crcr
               #1\crcr}}
\def\overrightarrow{\mathpalette\overrightarrow@}
\def\overrightarrow@#1#2{\vbox{\ialign{$##$\cr
    #1{-}\mkern-6mu\cleaders\hbox{$#1\mkern-2mu{-}\mkern-2mu$}\hfill
     \mkern-6mu{\to}\cr
     \noalign{\kern -1\p@\nointerlineskip}
     \hfil#1#2\hfil\cr}}}
\def\overleftarrow{\mathpalette\overleftarrow@}
\def\overleftarrow@#1#2{\vbox{\ialign{$##$\cr
     #1{\leftarrow}\mkern-6mu\cleaders\hbox{$#1\mkern-2mu{-}\mkern-2mu$}\hfill
      \mkern-6mu{-}\cr
     \noalign{\kern -1\p@\nointerlineskip}
     \hfil#1#2\hfil\cr}}}
\def\overleftrightarrow{\mathpalette\overleftrightarrow@}
\def\overleftrightarrow@#1#2{\vbox{\ialign{$##$\cr
     #1{\leftarrow}\mkern-6mu\cleaders\hbox{$#1\mkern-2mu{-}\mkern-2mu$}\hfill
       \mkern-6mu{\to}\cr
    \noalign{\kern -1\p@\nointerlineskip}
      \hfil#1#2\hfil\cr}}}
\def\underrightarrow{\mathpalette\underrightarrow@}
\def\underrightarrow@#1#2{\vtop{\ialign{$##$\cr
    \hfil#1#2\hfil\cr
     \noalign{\kern -1\p@\nointerlineskip}
    #1{-}\mkern-6mu\cleaders\hbox{$#1\mkern-2mu{-}\mkern-2mu$}\hfill
     \mkern-6mu{\to}\cr}}}
\def\underleftarrow{\mathpalette\underleftarrow@}
\def\underleftarrow@#1#2{\vtop{\ialign{$##$\cr
     \hfil#1#2\hfil\cr
     \noalign{\kern -1\p@\nointerlineskip}
     #1{\leftarrow}\mkern-6mu\cleaders\hbox{$#1\mkern-2mu{-}\mkern-2mu$}\hfill
      \mkern-6mu{-}\cr}}}
\def\underleftrightarrow{\mathpalette\underleftrightarrow@}
\def\underleftrightarrow@#1#2{\vtop{\ialign{$##$\cr
      \hfil#1#2\hfil\cr
    \noalign{\kern -1\p@\nointerlineskip}
     #1{\leftarrow}\mkern-6mu\cleaders\hbox{$#1\mkern-2mu{-}\mkern-2mu$}\hfill
       \mkern-6mu{\to}\cr}}}
\def\sqrt#1{\radical"270370 {#1}}
\def\dots{\relax\ifmmode\let\next=\ldots\else\let\next=\tdots@\fi\next}
\def\tdots@{\unskip\ \tdots@@}
\def\tdots@@{\futurelet\next\tdots@@@}
\def\tdots@@@{$\mathinner{\ldotp\ldotp\ldotp}\,
   \ifx\next,$\else
   \ifx\next.\,$\else
   \ifx\next;\,$\else
   \ifx\next:\,$\else
   \ifx\next?\,$\else
   \ifx\next!\,$\else
   $ \fi\fi\fi\fi\fi\fi}
\def\text{\relax\ifmmode\let\next=\text@\else\let\next=\text@@\fi\next}
\def\text@@#1{\hbox{#1}}
\def\text@#1{\mathchoice
 {\hbox{\everymath{\displaystyle}\def\textfonti{\the\textfont1 }%
    \def\textfontii{\the\textfont2 }\textdef@@ T#1}}
 {\hbox{\everymath{\textstyle}\def\textfonti{\the\textfont1 }%
    \def\textfontii{\the\textfont2 }\textdef@@ T#1}}
 {\hbox{\everymath{\scriptstyle}\def\textfonti{\the\scriptfont1 }%
   \def\textfontii{\the\scriptfont2 }\textdef@@ S\rm#1}}
 {\hbox{\everymath{\scriptscriptstyle}\def\textfonti{\the\scriptscriptfont1 }%
   \def\textfontii{\the\scriptscriptfont2 }\textdef@@ s\rm#1}}}
\def\textdef@@#1{\textdef@#1\rm \textdef@#1\bf
   \textdef@#1\sl \textdef@#1\it}

\def\textdef@#1#2{\def\next{\csname\expandafter\eat@\string#2fam\endcsname}%
\if S#1\edef#2{\the\scriptfont\next\relax}%
 \else\if s#1\edef#2{\the\scriptscriptfont\next\relax}%
 \else\edef#2{\the\textfont\next\relax}\fi\fi}
\scriptfont\itfam=\tenit \scriptscriptfont\itfam=\tenit
\scriptfont\slfam=\tensl \scriptscriptfont\slfam=\tensl
\mathcode`\0="0030
\mathcode`\1="0031
\mathcode`\2="0032
\mathcode`\3="0033
\mathcode`\4="0034
\mathcode`\5="0035
\mathcode`\6="0036
\mathcode`\7="0037
\mathcode`\8="0038
\mathcode`\9="0039
\def\Cal{\relax\ifmmode\let\next=\Cal@\else
     \def\next{\errmessage{Use \string\Cal\space only in math mode}}\fi\next}
\def\Cal@#1{{\fam2 #1}}
\def\bold{\relax\ifmmode\let\next=\bold@\else
   \def\next{\errmessage{Use \string\bold\space only in math
      mode}}\fi\next}\def\bold@#1{{\fam\bffam #1}}
\mathchardef\Gamma="0000
\mathchardef\Delta="0001
\mathchardef\Theta="0002
\mathchardef\Lambda="0003
\mathchardef\Xi="0004
\mathchardef\Pi="0005
\mathchardef\Sigma="0006
\mathchardef\Upsilon="0007
\mathchardef\Phi="0008
\mathchardef\Psi="0009
\mathchardef\Omega="000A
\mathchardef\varGamma="0100
\mathchardef\varDelta="0101
\mathchardef\varTheta="0102
\mathchardef\varLambda="0103
\mathchardef\varXi="0104
\mathchardef\varPi="0105
\mathchardef\varSigma="0106
\mathchardef\varUpsilon="0107
\mathchardef\varPhi="0108
\mathchardef\varPsi="0109
\mathchardef\varOmega="010A
\font\dummyft@=dummy
\fontdimen1 \dummyft@=\z@
\fontdimen2 \dummyft@=\z@
\fontdimen3 \dummyft@=\z@
\fontdimen4 \dummyft@=\z@
\fontdimen5 \dummyft@=\z@
\fontdimen6 \dummyft@=\z@
\fontdimen7 \dummyft@=\z@
\fontdimen8 \dummyft@=\z@
\fontdimen9 \dummyft@=\z@
\fontdimen10 \dummyft@=\z@
\fontdimen11 \dummyft@=\z@
\fontdimen12 \dummyft@=\z@
\fontdimen13 \dummyft@=\z@
\fontdimen14 \dummyft@=\z@
\fontdimen15 \dummyft@=\z@
\fontdimen16 \dummyft@=\z@
\fontdimen17 \dummyft@=\z@
\fontdimen18 \dummyft@=\z@
\fontdimen19 \dummyft@=\z@
\fontdimen20 \dummyft@=\z@
\fontdimen21 \dummyft@=\z@
\fontdimen22 \dummyft@=\z@
\def\fontlist@{\\{\tenrm}\\{\sevenrm}\\{\fiverm}\\{\teni}\\{\seveni}%
 \\{\fivei}\\{\tensy}\\{\sevensy}\\{\fivesy}\\{\tenex}\\{\tenbf}\\{\sevenbf}%
 \\{\fivebf}\\{\tensl}\\{\tenit}\\{\tensmc}}
\def\dodummy@{{\def\\##1{\global\let##1=\dummyft@}\fontlist@}}
\newif\ifsyntax@
\newcount\countxviii@
\def\newtoks@{\alloc@5\toks\toksdef\@cclvi}
\def\nopages@{\output={\setbox\z@=\box\@cclv \deadcycles=\z@}\newtoks@\output}
\def\syntax{\syntax@true\dodummy@\countxviii@=\count18
\loop \ifnum\countxviii@ > \z@ \textfont\countxviii@=\dummyft@
   \scriptfont\countxviii@=\dummyft@ \scriptscriptfont\countxviii@=\dummyft@
     \advance\countxviii@ by-\@ne\repeat
\dummyft@\tracinglostchars=\z@
  \nopages@\frenchspacing\hbadness=\@M}
\def\magstep#1{\ifcase#1 1000\or
 1200\or 1440\or 1728\or 2074\or 2488\or
 \errmessage{\string\magstep\space only works up to 5}\fi\relax}
{\lccode`\2=`\p \lccode`\3=`\t
 \lowercase{\gdef\tru@#123{#1truept}}}

\def\scaletype#1{\mag=#1\relax
 \hsize=\expandafter\tru@\the\hsize
 \vsize=\expandafter\tru@\the\vsize
 \dimen\footins=\expandafter\tru@\the\dimen\footins}

\def\scalefont#1#2\andcallit#3{\edef\font@{\the\font}#1\font#3=
  \fontname\font\space scaled #2\relax\font@}
\def\Mag@#1#2{\ifdim#1<1pt\multiply#1 #2\relax\divide#1 1000 \else
  \ifdim#1<10pt\divide#1 10 \multiply#1 #2\relax\divide#1 100\else
  \divide#1 100 \multiply#1 #2\relax\divide#1 10 \fi\fi}
\def\scalelinespacing#1{\Mag@\baselineskip{#1}\Mag@\lineskip{#1}%
  \Mag@\lineskiplimit{#1}}
\def\wlog#1{\immediate\write-1{#1}}
\catcode`\@=\active
%
%
\magnification=\magstep1
\TagsOnRight

{\nopagenumbers
\headline{\rightline{ CERN-TH.7364/94}}
\vglue 0.9cm
\centerline{ TWO-LOOP QUARK SELF-ENERGY IN A NEW FORMALISM}
\vglue 0.1cm
\centerline{ (I) OVERLAPPING DIVERGENCES }
\vglue 1.1cm
\centerline{ GEORGE LEIBBRANDT\footnote {
 Department of Mathematics and Statistics, University of Guelph,
 Guelph, Ontario, N1G \hglue 5pt 2W1, Canada.
 \hglue 5pt E-mail address:  gleibbra\@msnet\.mathstat\.uoguelph\.ca }}
\vglue 0.1cm
\centerline{ Theoretical Physics Division, CERN, CH-1211 Geneva 23}
\vglue 0.1cm
\centerline{ and}
\vglue 0.1cm
\centerline{ JIMMY WILLIAMS\footnote {
 Department of Physics, University of Guelph,
 Guelph, Ontario, N1G 2W1, Canada. \newline
 \hglue 5pt E-mail address:  williams\@physics\.uoguelph\.ca}}
\vglue 1.1cm
\centerline{ ABSTRACT}
\vglue 0.3cm
{\baselineskip=16pt \noindent \narrower
A new integration technique for multi-loop Feynman integrals, called the
{\it matrix method}, is developed and then applied to the divergent part of
the overlapping two-loop quark self-energy function $\,i\Sigma\,$ in the
light-cone gauge $\ n\!\cdot\!A^a(x)=0,\ n^2=0$.  It is shown that the
coefficient of the double-pole term is strictly local, even off mass-shell,
while the coefficient of the single-pole term contains local as well as
nonlocal parts.  On mass-shell, the single-pole part is local, of course.
It is worth noting that the original overlapping self-energy integral reduces
eventually to 10 covariant and 38 noncovariant-gauge integrals.  We were able
to verify explicitly that the {\it divergent parts} of the 10 double
covariant-gauge integrals agreed precisely with those currently used to
calculate radiative corrections in the Standard Model.

   Our new technique is amazingly powerful, being applicable to massive and
massless integrals alike, and capable of handling both covariant-gauge
integrals and the more difficult noncovariant-gauge integrals.  Perhaps the
most important feature of the matrix method is the ability to execute the
$4\omega$-dimensional momentum integrations in a single operation, exactly
and in analytic form.  The method works equally well for other axial-type
gauges, notably the temporal gauge ($n^2>0$) and the pure axial gauge
($n^2<0$).

}
\newpage}
\baselineskip=20pt \count0=1
{\narrower

\leftline{\bf 1. Introduction}
\vskip 15pt
       Traditional perturbation theory with its emphasis on Feynman diagrams
continues to play a central role in quantum field theory.  The success of
the perturbative approach hinges decisively on the accurate computation
of multi-loop Feynman integrals.  Of course, there exists a vast variety of
such integrals, but here we shall only distinguish between covariant-gauge
Feynman integrals and noncovariant-gauge integrals.  As the name implies,
covariant-gauge Feynman integrals occur in theories quantized in a
covariant gauge, such as the Fermi gauge or the unitary gauge.
Noncovariant-gauge integrals, by contrast, arise whenever a noncovariant
gauge is implemented, such as the powerful light-cone gauge [1] or the
infamous Coulomb gauge [2].  While both types of integrals are known to
require patience and a healthy respect for detail, it is also true that
noncovariant integrals remain as popular as skunks at a garden party, as
seen from the small fraction of higher-loop calculations in QCD and
Yang-Mills theory [3-8].

\medpagebreak
According to the literature, multi-loop integrals, such as
$\,\int\! d^{2\omega}q\int\! d^{2\omega}k\,f(q,k)$, have been evaluated almost
exclusively by the {\it nested method} [9-14], in which the four-momentum
integrations are carried out sequentially (initial exponential parametrization
of the propagators is assumed).  For instance, for the double integral
mentioned above, one first integrates over $\,d^{2\omega} k$, then over
$\,d^{2\omega} q$ (or conversely).  Dimensional regularization will be used
throughout this paper, with $ 2\omega$ denoting the dimensionality of
complex space-time.  We have no intention of reviewing here the dominant
characteristics and various idiosyncrasies of the nested approach, except to
say that it has been used and abused with varying degrees of success.

\medpagebreak
Instead, we should like to propose an alternative procedure
to the nested method, called the {\it matrix integration technique}, in
which the two momentum integrals in
$\,\int d^{2\omega} q \int d^{2\omega} k\,f(q,k)\,$
are integrated over $4\omega$-dimensional space in a {\it single operation}.
Perhaps the most appealing feature of this method is the ability to perform
the momentum integration exactly and in closed form, thereby guaranteeing
from the outset a certain amount of what might be called ``calculational
streamlining''.  In the course of our investigation we shall encounter several
examples of this ``streamlining''.  It turns out that the matrix technique
works for covariant and noncovariant gauges alike, and regardless whether
the integrals are massive or massless.

\medpagebreak
The purpose of our integration program may now be stated as follows:
\item{1.\,} To develop for two-loop integrals an alternative approach to the
nested method, called the {\it matrix integration technique}.
\item{2.\,} To apply this technique to the complete two-loop quark self-energy
function in the light-cone gauge $\,n\!\cdot\! A^a(x)=0,\,n^2=0$.
\item{3.\,} To derive, if possible, the necessary counterterms and use these
to renormalize the theory to two-loop order.
\item{4.\,} To extend the matrix technique to other axial gauges, notably
to the temporal gauge defined by $\,n\!\cdot\! A^a(x)=0,\,n^2>0$,
and to the pure axial gauge  $\,n\!\cdot\! A^a(x)=0,\,n^2<0$.

\smallpagebreak \flushpar
As indicated above, the testing ground for our matrix technique will be the
quark self-energy function to two loops, depicted graphically in Figures
1 and 2.  For pedagogical reasons, we have decided to report our results
in two separate papers.  In paper I, we shall discuss various mathematical
tools and then apply these to compute the divergent portion of the
{\it overlapping} quark self-energy (Fig.\,1).  The remaining two-loop
graphs, including the non-trivial rainbow diagram (Fig.\,2(a)),
will be treated in paper II, where we shall also study the counterterms
required for renormalization.

\medpagebreak
The plan of paper I is as follows.  In Section 2 we define the structure of
the overlapping quark self-energy function and review some standard formulas
needed for one-loop calculations.  In Section 3, we summarize the main
features of the matrix method, including reduction of the integrand to
Gaussian form and the evaluation of $4\omega$-dimensional
Gaussian integrals.  The parameter integrations are carried out in Sections
4 and 5.  But in order to prepare the reader for certain technical subtleties,
we shall first make a short detour to introduce some terminology.

\medpagebreak
For each factor in the denominator of the integrand of
$\,\int d^{2\omega} q \int d^{2\omega} k\,f(q,k)$, there will be a
\it Schwinger parameter\rm, $\alpha_j$ say ($j=1,2,\,\dots$), with an
infinite domain of integration; i.e., $\alpha_j\,\epsilon\ [0,\infty]$.
For the overlapping diagram discussed in this article, we
transform the set of parameters $\{\alpha_j\}$ to a more
``user-friendly'' set $S$, containing two types of parameters:
one \it type I \rm parameter ($A$), with an infinite domain, and up to
six \it type II \rm parameters ($\lambda,\beta,G,b,h,a$), with finite
domains.  We shall demonstrate in Section 4 that integration over the
lone type I parameter leads to a simple pole, and in Section 5 that
one more simple pole (and hence a double pole overall) emerges from
integration over one of the finite type II parameters.  The paper
concludes in Section 6 with a short discussion.

\vskip 30pt

\leftline {\bf 2. Basic Tools }
\leftline {\bf \ \ (a) Notation and review of the light-cone prescription }
\vskip 15pt
In Yang-Mills theory, the light-cone gauge is characterized by
  \flushpar {\hsize=371pt \vskip -12pt $$
  n^\mu A^a_\mu(x)=0,\qquad n^2=0,\qquad \mu = 0,1,2,3, \tag 2.1
  $$ \vskip 2pt } \flushpar
where $\,n_\mu =(n_0, \vec n)\,$ defines a fixed axis in four-space [1].  We
use a ($+,-,-,-$) metric and employ dimensional regularization in a space-time
of $2\omega$ dimensions.  The relevant SU$(3)_c\,$ color Lagrangian density
reads [15]
  \flushpar {\hsize=371pt \vskip -20pt $$ \align
  \Cal L\ &=\ \Cal L_{int} - \lim_{\lambda\to 0} \tfrac1{2\lambda}
    (n\!\cdot\!A^a)^2,\qquad n^2=0, \tag 2.2
  \\ \vspace{12pt}
  \Cal L_{int}\ &= \ -\tfrac14 F^a_{\mu\nu} F^{a\mu\nu} + \sum_k \,i
    \overline\psi^k_\alpha
    (\gamma_\mu D^\mu_{\alpha\beta}+m_k \delta_{\alpha\beta})
    \psi^k_\beta\,, \tag 2.3 \endalign
  $$ \vskip 2pt } \flushpar
where $\,\psi_\alpha$ and $A^a_\mu\,$ represent fermion and gluon fields,
respectively, $\,\gamma_\mu\,$ are $4\times 4$ Dirac gamma-matrices,
$\,a = 1,2,\,\dots\,,8\,$ is the group index,
$\,\alpha,\beta\,$ are color indices, $\,\lambda\,$ is the gauge parameter,
$\,k=u,d,\,\dots\,$ is the quark flavor index, and $\,m_k\,$ are quark rest
masses.  Moreover,
  \flushpar {\hsize=371pt \vskip -12pt $$
  F^a_{\mu\nu}\ =\ \partial_\mu A^a_\nu - \partial_\nu A^a_\mu
    +gf^{abc}A^b_\mu A^c_\nu  \tag 2.4a
  $$ \vskip 2pt } \flushpar
is the field strength ($g$ is the QCD coupling constant), and
  \flushpar {\hsize=371pt \vskip -12pt $$
  D_{\alpha\beta,\mu}\ =\ \delta_{\alpha\beta}\partial_\mu-
    igT^a_{\alpha\beta} A^a_\mu  \tag 2.4b
  $$ \vskip 2pt } \flushpar
denotes the covariant derivative ($\,\partial_\mu \equiv
\partial/\partial x^\mu). \ \ T^a_{\alpha\beta}\,$ are the generators of
the gauge group SU$(3)$ which obey the commutation relations
  \flushpar {\hsize=371pt \vskip -12pt $$
  [T^a,T^b]\ =\ if^{abc}T^c, \tag 2.4c
  $$ \vskip 2pt } \flushpar
$f^{abc}\,$ being antisymmetric structure constants.  In the light-cone gauge,
the Lagrangian density (2.2) leads to the gluon propagator
  \flushpar {\hsize=371pt \vskip -10pt $$
  G^{ab}_{\mu\nu}(q) \ = \ \frac{-i\,\delta^{ab}}{q^2+i\epsilon}
    \left(g_{\mu\nu}-\frac{q_\mu n_\nu+q_\nu n_\mu}{q\!\cdot\!n} \right),
  \qquad \epsilon > 0. \tag 2.5
  $$ \vskip 6pt } \flushpar
The spurious poles of the factor $(q\!\cdot\!n)^{-1}$ will be handled by the
$\,n^*_\mu$-prescription [16-17]:
  \flushpar {\hsize=371pt \vskip -24pt $$
  \frac1{q\!\cdot\!n}\ =\ \lim_{{\dsize\epsilon}\to 0}
    \,\frac{q\!\cdot\!n^*}{q\!\cdot\!n\ q\!\cdot\!n^*+i\epsilon}\,,
  \qquad \epsilon > 0, \tag 2.6
  $$ \vskip 6pt } \flushpar
with $\,n_\mu=(n_0,\vec n)$ and $\,n^*_\mu=(n_0,-\vec n)$.

\pagebreak
\leftline {\bf \ \ (b) The overlapping quark self-energy }
\vskip 15pt
Application of the Feynman rules for QCD, with the quark-quark-gluon
vertex factor $\,igT^a_{\alpha\beta} \gamma^\mu $, the quark propagator
$\,i/(q\!\!\!/-m)$, and the gluon propagator (2.5), yields the following
expression for the two-loop overlapping quark self-energy function in the
light-cone gauge (Fig\. 1):
  \flushpar {\hsize=371pt \vskip -16pt $$
  \align
  & I_0 = \ ig^4 T^b_{DA} T^a_{CD} T^b_{BC} T^a_{AB} I, \tag 2.7
  \\ \vspace{16pt}
  & I = \,\int_M \int_M \frac{d^4 q d^4 k}{(2\pi)^8}
    \frac{\gamma^\mu\ (p\!\!\!/ - q\!\!\!/ +m)
       \ \gamma^\tau\ (p\!\!\!/ - k\!\!\!/ - q\!\!\!/ +m)
       \ \gamma^\nu \ (p\!\!\!/ - k\!\!\!/ +m)\ \gamma^\sigma}
      {q^2\,[(p-q)^2-m^2]\ [(p-k-q)^2-m^2]\ [(p-k)^2-m^2]\ k^2}
  \\ \vspace{12pt}
  & \qquad\qquad\qquad\qquad \times
    \left(g_{\mu\nu}-\frac{n_\mu q_\nu+n_\nu q_\mu}{n \!\cdot\! q}\right)
    \left(g_{\sigma\tau}-\frac{n_\sigma k_\tau+n_\tau k_\sigma}
                              {n \!\cdot\! k}\right), \tag 2.8 \endalign
  $$ \vskip 6pt } \flushpar
where $\,m\,$ is the quark mass, $\,p\,$ the quark 4-momentum,
$\int_M$ denotes integration over Minkowski
space, and where the five $\,i\epsilon\,$ terms have been omitted.
Evaluation of the integral $I$ by means of the matrix method is the
main objective of this article.

\medpagebreak
To begin with, we observe that $I$ in Eq\. (2.8) diverges at the following
three types of limits:
\itemitem{(i) }$|q|$ and/or $|k|\ \to\infty\,$;
\itemitem{(ii)\ }$q$ and/or $k\ \to$ zeros of the quadratic factors
in the denominator of Eq\. (2.8);
\itemitem{(iii)\ }$q\!\cdot\!n\,\to\,0\ \ $
and/or $\ \ k\!\cdot\!n\,\to\,0\,$.

\smallpagebreak \flushpar
The first two types of divergence are handled by dimensional regularization
and Wick rotation.  The third type of divergence (iii) is symptomatic of
axial-type gauges, and requires application of the $\,n^*_\mu$-prescription
(2.6) to the spurious poles of $(q\!\cdot\!n)^{-1}$ and $(k\!\cdot\!n)^{-1}$.
Since the $\,n^*_\mu$-prescription permits a Wick rotation $(iq_4=q_0,\,
ik_4=k_0,\,ip_4=p_0)$, we may \pagebreak
Wick-rotate the entire integrand from Minkowski space to Euclidean space.

\medpagebreak
Next we apply exponential parametrization to every quadratic factor $F_j$
\linebreak $(j=1,2,\,\dots)$ in the denominator of the rotated integrand:
  \flushpar {\hsize=371pt \vskip -12pt $$
  \frac1{F_j}\ =\ \int_0^\infty d\alpha_j\ \exp{(-\alpha_j F_j)}\,,
  \qquad F_j>0\,, \tag 2.9
  $$ \vskip 6pt } \flushpar
thereby replacing the product of quadratic factors in the denominator by
a {\it sum} of quadratic terms in the exponent.  Completing the square
in this exponent, we can then integrate over the momentum variables by
means of the generalized Gaussian formula:
  \flushpar {\hsize=371pt \vskip -12pt $$
  \int d^{2\omega}q\ \exp{(-A q^2)}\,\ =\,\ \left(\frac\pi A\right)^\omega,
  \qquad A\,>\,0.  \tag 2.10 $$ }

\vskip 10pt
\leftline {\bf \ \ (c) Two-loop integrations:  the traditional approach }
\vskip 15pt
    Although the methods described in Section 2(b) are applicable to any
multi-loop integral, the ever-increasing technical difficulties have
drastically restricted the number of explicit computations.  Here is a
brief look at some typical problems, encountered already in double
integrals such as $\,\int d^{2\omega} q \int d^{2\omega} k\,f(q,k)$.

\medpagebreak
   Two-loop integrals have traditionally been computed by the so-called \it
nested method\rm.  As alluded to in the \underbar{Introduction}, the basic
idea behind the nested method is to integrate over the loop momenta one
loop at a time, i.e., to integrate over $\,k\,$ as completely as possible,
before attempting integration over $\,q$.  In many two-loop cases the
$\,k$-integral is so complicated, however, that we cannot fully compute
it, unless we express the result in the form of a series, such as the
Laurent series $\,\sum_j\ (\omega-2)^j\ T_j(n,p,q;m)$.
Each of the functions $T_j$ must then be multiplied by the
remaining $\,q$-dependent factors in the original integrand, and the
resulting expressions integrated over $2\omega$-dimensional $\,q$-space.
One hopes,
of course, that only the first few (i.e., small $\,j$) terms of
the Laurent series will contribute to the pole parts of the final result.
Unfortunately, no such luck prevails in general, as may be seen from the
following typical, albeit simplified, example:
  \flushpar {\hsize=371pt \vskip -12pt $$
  \int_1^\infty q^{1-\omega}\ q^{2\omega-4}\ dq\ \to
  \ \sum_{j=0}^\infty \left( \frac{(2\omega-4)^j}{j!}
  \int_1^\infty q^{1-\omega}\ (\ln q)^j\ dq \right). \tag 2.11
  $$ \vskip 6pt } \flushpar
Integration of the $\,j$-th term on the right-hand side by parts $\,j\,$ times
yields $\sum_j\,2^j$, whereas direct evaluation of the left-hand side yields
$(2-\omega)^{-1}$.  The inconsistency stems from the fact that the
exponential series for $\,q^{2\omega-4}\,$ does \it not converge
uniformly \rm as $|q|\to\infty$.

\medpagebreak
In order to improve the behaviour for large values of $|q|$, we could try to
express the result of the $\,k$-integration as a series in powers of $\,1/q^2$.
One may always obtain such a series by applying the binomial formula
to the integrand \it before \rm integrating over the last one or two Schwinger
parameters associated with the $\,k$-integration.  On the surface, this
strategy looks promising, but on closer inspection we sometimes find that the
resulting series \it fails to converge \rm for certain combinations of
values of $\,q\,$ and the unintegrated Schwinger parameters.  Some of the UV
divergences of the integral $I$, for instance, occur precisely in regions
where the binomial series diverges.

\medpagebreak
These simple examples serve as a potent reminder that (1) the divergent
terms for overlapping loops do not arise merely from the
divergent and finite terms ($j\leq 0$) of the individual loops, and that (2)
care must be exercised in choosing series which converge uniformly over
the entire region of integration.  We shall see in Sections 4 and 5 that
point (2) is also crucial in the context of the matrix method, where
similar convergence issues arise, albeit in vastly simplified form.

\pagebreak
\leftline {\bf 3. The Matrix Method }
\leftline {\bf \ \ (a) Basics }
\vskip 12pt
   In this section we propose an alternative procedure to the nested
method, called the matrix integration technique, or \it matrix method \rm
for short.  In the matrix method we regard $q$-space and $k$-space as
subspaces of a single $4\omega$-dimensional momentum space and proceed
as in the case of one-loop integrals.  In other words, we Wick-rotate
$\,q_0\,$ and $\,k_0$, use the $\,n^*_\mu$-prescription (2.6) for
$(q\!\cdot\!n)^{-1}$ and $(k\!\cdot\!n)^{-1}$, and apply the
exponential parametrization (2.9) to all denominator factors.  The two-loop
integral $I$ in Eq\. (2.8) for the overlapping diagram in Figure 1 then
assumes the form:
  \flushpar {\hsize=371pt \vskip -16pt $$
  \int_E\! d^{2\omega}q \int_E\! d^{2\omega}k \ f(q,k)
  \,= \,\int_0^\infty\! d\alpha_1 \int_0^\infty\! d\alpha_2\,\dots\
  J[P(q,k)] \qquad \equiv\,I_E[f]\,, \tag 3.1
  $$ \vskip 5pt } \flushpar
where:
  \flushpar {\hsize=371pt \vskip -20pt $$ \align
  & f(q,k)\,=\ \frac{i^2(-1)^4\ \gamma^\mu\ (p\!\!\!/ - q\!\!\!/ -m)
             \ \gamma^\tau\ (p\!\!\!/ - k\!\!\!/ - q\!\!\!/ -m)
             \ \gamma^\nu \ (p\!\!\!/ - k\!\!\!/ -m)\ \gamma^\sigma}
    {(2\pi)^{4\omega}\ q^2\,[(p-q)^2+m^2]\ [(p-k-q)^2+m^2]\ [(p-k)^2+m^2]\ k^2}
  \\ \vspace{10pt}
  & \quad \times
    \frac{(n\!\cdot\!q\delta_{\mu\nu}-n_\mu q_\nu-n_\nu q_\mu)}{n \!\cdot\! q}
    \frac{(n\!\cdot\!k\delta_{\sigma\tau}-n_\sigma k_\tau-n_\tau k_\sigma)}
        {n \!\cdot\! k}          \qquad \text{(in this case);} \tag 3.2
  \\ \vspace{20pt}
  & J[P(q,k)]\ \equiv \int_E d^{2\omega}q \int_E d^{2\omega}k \ P(q,k)
    \ \exp(-\,\bold z^\top\bold M\bold z\ +\ 2\bold B^\top\bold z\ -\ C)\,;
    \tag 3.3
  \\ \vspace{15pt}
  & P(q,k)\ \equiv \ \text{the numerator of $f$, multiplied by
     $\,-q\!\cdot\! n^*/n_0^2\,$ and/or}
  \\ \vspace{-4pt}
  & \ \ \ \qquad \qquad -k\!\cdot\! n^*/n_0^2\,\ \text{
     from the $\,n^*_\mu$-prescription (as applicable);} \endalign
  $$ \vskip -2pt } \flushpar
$\,\bold z\ \equiv\ (k_4,\,q_4,\,k_3,\,q_3,\,\dots\,)^\top
=\,$ a $4\omega$-component column-vector (${}^\top\equiv$ transpose);
$\,\bold M\,$ is a $4\omega\times 4\omega\,$ matrix;
$\,\bold B\,$ is a $4\omega$-component column-vector;
$\,C\ $ is a scalar; and $\int_E$ denotes integration over Euclidean space.
Since $\bold M_{ij}$ is the coefficient of $\,z_i z_j\,$ in the exponent of
\pagebreak
the integrand in Eq\. (3.3), we may always symmetrize $\bold M$ \,(``$z_i$''
refers to the $i^{\text{th}}$ component of $\bold z$).

In the case of the integral $I$, with $\,f\,$ given by Eq\. (3.2),
let us take $\,\alpha_1,\,\dots\,,\alpha_7\,$ to be the parameters
corresponding to the factors $\,n\!\cdot\! q,\ q^2,\ [(p-q)^2+m^2],
\ [(p-k-q)^2+m^2],\ [(p-k)^2+m^2],\ k^2,\,$ and $\,n\!\cdot\! k,\,$
respectively.  We then find that
$\bold M,\,\bold B,$ and $C$ take particularly simple forms when
we change the variables of integration from $\{\alpha_1,\,\dots\,,\alpha_7\}$
to the more ``user-friendly'' set $S=\{A;\,\lambda,\beta,G,b,h,a\}$, where
  \flushpar {\hsize=371pt \vskip -8pt $$  \left. \matrix
  A \equiv\ \alpha_1+\alpha_2+\alpha_3+2\alpha_4+\alpha_5+\alpha_6+\alpha_7\,,
  \\ \vspace{15pt}
  G \equiv \ \alpha_4/A\,,\qquad b \equiv \ (\alpha_4+\alpha_5)/A\,,
  \\ \vspace{12pt}
  \beta \equiv \ (\alpha_3+\alpha_4)/A\,,\qquad
  h \equiv \ (\alpha_4+\alpha_5+\alpha_6)/A\,,\\ \vspace{12pt}
  \lambda\equiv \ (\alpha_2+\alpha_3+\alpha_4)/A\,,\qquad
  a \equiv \ (\alpha_4+\alpha_5+\alpha_6+\alpha_7)/A\,.
  \endmatrix \qquad \right\} \tag 3.4
  $$ \vskip 12pt } \flushpar
As explained in the \underbar{Introduction}, $A$ is a type I parameter with
an infinite domain, while $\,\lambda,\beta,G,b,h,a\,$ are type II parameters
with finite domains.  In terms of these parameters,
  \flushpar {\hsize=371pt \vskip -10pt $$  \left. \matrix
  \bold M = A\left[ \matrix
  a & G &   &   &  &    &  & \\ G &\,1-a\! & &     &   &         &   &
  \\ \vspace{4pt}
   &   & a & G &  &    &  & \\   &     & G &\,1-a\! & &         &   &
  \\ \vspace{4pt}
   &  &  &  & h\,&\,G &  & \\   &     &   &     & G & \lambda &   &
  \\ \vspace{4pt}
   &  &  &  &  &  & h\,&\,G \\   &     &   &     &   &   & G & \lambda
  \endmatrix \right], \quad \bold B = A\left[ \matrix
  bp_4 \\ \beta p_4 \\ \vspace{4pt} bp_3 \\ \beta p_3 \\ \vspace{4pt}
  bp_2 \\ \beta p_2 \\ \vspace{4pt} bp_1 \\ \beta p_1 \endmatrix \right],
  \\ \vspace{12pt}
  C = A\,(b+\beta -G)\,( p^2+m^2)\,,\qquad\qquad\qquad\qquad
              \,\ \qquad\qquad\qquad\qquad  \endmatrix \ \right\} \tag 3.5
  $$ \vskip 12pt } \flushpar
where $\,n_1=n_2=0\,$ for simplicity.  Note that the
lower half of $\bold M$ has $\ 2\omega-2\,$ pairs
of rows, and the right-hand half $\ 2\omega-2\,$ pairs of columns, so that
  \flushpar {\hsize=371pt \vskip -15pt $$ \align
  \sqrt{\det(\bold M)}\ &=\ A^{2\omega} D_\parallel\,D_\perp^{\omega-1}\,,
  \tag 3.6 \\ \vspace{2pt}
  \text{where}\qquad D_\parallel &\equiv a(1-a)-G^2\,,\qquad
              D_\perp \equiv \lambda h-G^2\,. \tag 3.7 \endalign
  $$ \vskip -2pt }

\pagebreak
\leftline {\bf \ \ (b) Momentum Integration }
\vskip 15pt
Our first major task is to evaluate the momentum-space integral $J[P]$
in Eq\. (3.3), $P$ being a polynomial in the components of $\,q\,$ and $\,k$.
Differentiating Eq\. (3.3) partially with respect to $B_i\,$, we obtain
  \flushpar {\hsize=371pt \vskip -10pt $$
  \frac{\partial J[P]}{\partial B_i} \ = \ 2\,J[z_iP]. \tag 3.8
  $$ \vskip 2pt } \flushpar
Once $J[1]$ is known, we can derive $J[P]$ for any polynomial $P$ by repeated
application of formula (3.8).

\smallpagebreak
   To evaluate $J[1]$, we first diagonalize the quadratic form in the
exponent of the integrand in Eq\. (3.3).  Since $\bold M$ is symmetric, there
exists a matrix $\bold R$ such that $\bold R^\top\bold R = \bold 1\,$, and
$\bold R\bold M\bold R^\top\,$ is diagonal.  Letting $\,\bold z =\bold R^\top
\bold y +\bold M^{-1}\bold B\,$ in Eq\. (3.3), and choosing $P=1$, we obtain
  \flushpar {\hsize=371pt \vskip -18pt $$
  J[1] \ =\ \exp{(\bold B^\top\bold M^{-1}\bold B \ - \ C)}
  \ \int_E\exp{(-\bold y^\top\bold R\bold M\bold R^\top\bold y)}
  \ d^{4\omega}\bold y\,. \tag 3.9
  $$ \vskip 2pt } \flushpar
With $\bold R\bold M\bold R^\top\,$ diagonal, the above
integral is just a product of one-dimensional Gaussian integrals, so that
  \flushpar {\hsize=371pt \vskip -10pt $$
  J[1] \ = \ \frac{\pi^{2\omega}\ \exp{(\bold B^\top\bold M^{-1}\bold B - C)}}
                  {\sqrt{\det(\bold M)}}\ , \tag 3.10
  $$ \vskip 6pt } \flushpar
$\det(\bold R)$ and $\det(\bold R^\top)$ having cancelled each other.
Applying formula (3.8) repeatedly to Eq\. (3.10), we get
  \flushpar {\hsize=371pt \vskip -20pt $$ \align
  J[z_i] \ &= \ m_i\,J[1]\,,\tag 3.11a \\
  J[z_i z_j] \ &= \ [m_i m_j+(\tfrac{\bold M^{-1}}2)_{ij}]\,J[1]\,,\tag 3.11b
  \\ \vspace{4pt}
  J[z_i z_j z_k] \ &= \ [m_i m_j m_k + m_i(\tfrac{\bold M^{-1}}2)_{jk} +
    m_j(\tfrac{\bold M^{-1}}2)_{ik} + m_k(\tfrac{\bold M^{-1}}2)_{ij}]\,J[1]\,,
    \ \ \ \ \ \tag 3.11c
  \\ &\vdots \endalign
  $$ \vskip 0pt } \flushpar
where $\ \bold m\equiv\bold M^{-1}\bold B$, and
$\ \partial m_i/\partial B_j\,=(\bold M^{-1})_{ij}\,$.

\medpagebreak
When $\bold M$ and $\bold B$ are given by Eqs\. (3.5), $D_\parallel$
and $D_\perp$ by Eqs\. (3.7), and $p_\parallel$ and $p_\perp$ by
  \flushpar {\hsize=371pt \vskip -25pt $$ \align
  &p_\parallel \,\equiv \,(0,\ 0,\ p_3,\,p_4) \ \ = \,\frac{n^*\!\!\cdot\! p
     \ n + n\!\cdot\! p \ n^*}{n^*\!\!\cdot\! n}\,,                \tag 3.12a
  \\ \vspace{6pt}
  &p_\perp \,\equiv \,(p_1,\,p_2,\ 0,\ 0) \ \ = \,p-p_\parallel\,, \tag 3.12b
  \\ \vspace{15pt} \text{we find}\quad
  &\qquad \bold m = (r_4,\,s_4,\,r_3,\,s_3,\,\dots\,)^\top, \tag 3.13
  \\ \vspace{15pt} \text{with} \qquad
  &r = \ r_\parallel + r_\perp \ =
    \ \left(\frac{(1-a)b-G\beta}{D_\parallel}\right) p_\parallel
    +\left(\frac{\lambda b-G\beta}{D_\perp}\right) p_\perp\,,\quad\tag 3.14a
  \\ \vspace{15pt}
  &s = \ s_\parallel + s_\perp \ =
    \ \left(\frac{a\beta-Gb}{D_\parallel}\right) p_\parallel
    +\left(\frac{h\beta-Gb}{D_\perp}\right) p_\perp \,. \tag 3.14b
  \endalign
  $$ \vskip 10pt } \flushpar
(For the sake of clarity, we have dropped the Lorentz indices on
$\,p_\parallel,\,p_\perp,\,n,\,n^*,$ $r,\,r_\parallel,\,r_\perp,\,s,\,
\,s_\parallel$, and $\,s_\perp$.)
Substituting from Eqs\. (3.5) and (3.13) into Eqs\. (3.10) and (3.11), and
exploiting the linearity of $J[P]$, we obtain the useful momentum-space
integrals given in Appendix A.

\bigpagebreak
The off-diagonal entries in $\bold M$ originate from the $\,q\!\cdot\! k\,$
term in the exponent of the integrand in Eq\. (3.3).  In the integral
$I$, this term comes, of course, from the parametrization of the
denominator factor $[(p-k-q)^2+m^2]$, which in turn comes
from the internal line shared by the two loops in Figure 1.
Note that in the \it absence \rm of such a shared line, the
integral $I$ would factor into two separate one-loop integrals.
By diagonalizing $\bold M$, it might appear that we have effectively
disentangled the overlapping loops, but that is not the case.  All we have
done is shift the problem to the parameter integrations.  To see this,
we observe that the integrands of
these integrations still contain the off-diagonal elements of $\bold M$
by virtue of the factors $\,\bold M^{-1}\,$ and $\,\sqrt{\det(\bold M)}\,$
in Eqs\. (3.10) and (3.11).  Parameter integrations will be discussed in
Sections 4 and 5.

\newpage
\leftline {\bf \ \ (c) Reduction of the integrand }
\vskip 15pt
   We observe in Eqs\. (3.11) that the complexity of $J$ increases
dramatically with the degree of its argument $P$, $P$ being the numerator of
the integrand $f$ in Eq\. (3.1) multiplied by $\,- q\!\cdot\! n^*/n_0^2\,$
and/or $\,- k\!\cdot\! n^*/n_0^2\,$ from the $n_\mu^*$-prescription if
applicable.  It is desirable, therefore, to reduce the degree of $P$
as much as possible before integration.  For $I$, the initial
degree of $P$ is 7, but we can reduce this number
to 3 by executing the following trivial operations before Wick-rotation:

\bigpagebreak
\item{1.} \ We completely expand the numerator of the integrand in Eq\. (2.8)
into a sum of products.  We then drop all terms which are shown by power
counting to be UV-convergent, since we are only interested in
the divergent terms.  We recall that the $n_\mu^*$-prescription
is consistent with power counting [1], and that Weinberg's Theorem
asserts that a two-loop integral is UV-convergent if the net power
of the integrand in each of $\,q,\,k$, and both
$\,q\,$ and $\,k\,$ together, is less than zero [18].  For
$I$, the result is that terms containing more than one factor of $m$
and/or $p\!\!\!/$ may be dropped.

\bigpagebreak
\item{2.} \ In the remaining terms, we rearrange the non-commuting Dirac
matrix factors with the help of the relation
$\,[\gamma^\mu,\gamma^\nu]_+ = 2g^{\mu\nu}$, so that
most terms with more than one matrix factor cancel with one another.
Some terms involving $\,q\!\!\!/ k\!\!\!/ n\!\!\!/\,$ remain,
but in these terms we may replace $\,q\!\!\!/ k\!\!\!/\,$ by
$\,\frac12(q\!\!\!/ k\!\!\!/ + k\!\!\!/ q\!\!\!/)=q\!\cdot\! k\,$ because
of the symmetry between $\,q\,$ and $\,k\,$ in Eq\. (2.8).   Also note that
$\,\gamma^\mu\gamma_\mu = 2\omega$, and that
$\,n^2=0\,$ in the light-cone gauge.

\bigpagebreak
\item{3.} \ In each of the remaining terms, we cancel as many factors
as possible with factors in the denominator of the integrand.  (In some
cases, identities such as $\mathbreak
\ 2q\!\cdot\! k \ =\ p^2-m^2-[(p-q)^2-m^2]+[(p-k-q)^2-m^2]-[(p-k)^2-m^2]\,,
\mathbreak
2p\!\cdot\! q \ =\ p^2-m^2-[(p-q)^2-m^2]+q^2\,,\,$ etc\. are helpful.)
The resulting integrands will have fewer denominator factors than were
present originally in Eq\. (2.8), but Eqs\. (3.5) to (3.14) are still valid,
provided we set the $\,\alpha\,$ parameters
which correspond to absent factors to zero in Eqs\. (3.4), and omit
the integrations over these parameters from Eq\. (3.1).
(We see from Eq\. (2.9) that this procedure is equivalent to inserting
factors of $\exp(0)$ in the integrand in Eq\. (3.3).  The
advantage of this strategy is obvious:  it allows us to use one common
expression for $J$ for all integrals with the same $P$, even though the
integrands of all these integrals will have different denominators.)

\bigpagebreak
\item{4.} \ Finally, we reduce the number of independent terms by
interchanging $\,q\,$ and $\,k\,$ in selected terms.

\bigpagebreak
The above four steps were carried out by means of a computer program
designed by one of the authors (J.W.).
Application of these steps, followed by Wick-rotation, transforms the
integral $I$ into the sum of 53 integrals, represented by the 53 non-blank
entries in Tables 1 to 6.  Each of the 53 integrals is equal to an entry in
one of the tables times the Euclidean-vector expression at the left-hand side
of its row of the table, divided by $\,(2\pi)^{4\omega}\,$ times the
denominator at the bottom of the table, and integrated over
$4\omega$-dimensional $\,qk$-space.

\medpagebreak
The symbols \,n, q, Q, ${}_\wedge$, K, and k \,at
the bottoms of the tables represent denominator factors in accordance with
the following scheme:

\vskip -12pt $$ \matrix \text{n } & \text{q} & \text{Q} & {}_\wedge
& \text{K} & \text{k} & \text{n } \\ \vspace{6pt}
n\!\cdot\! q & q^2 & [(p-q)^2+m^2] & [(p-k-q)^2+m^2] & [(p-k)^2+m^2] & k^2 &
n\!\cdot\! k \endmatrix
$$
\flushpar {\hsize=371pt \vskip -20pt $$ \tag 3.15 $$ \vskip 3pt }
\flushpar
Thus, an ``n'' on the left side of the denominator represents
$\,n\!\cdot\! q$, while an ``n'' on the right represents
$\,n\!\cdot\! k$.  For example, the entry $\,4-4\epsilon\,$ in the fifth row
and second-last column of Table 1 corresponds to the Euclidean-space integral
  \flushpar {\hsize=371pt \vskip -10pt $$
  \frac{(4-4\epsilon) \ n\!\cdot\!p} {(2\pi)^{4\omega}}
  \int_E \int_E  \,\frac  {q\!\!\!/ \ \ d^{2\omega}q\ d^{2\omega}k}
  {\,n\!\cdot\!q \ q^2\,[(p-q)^2+m^2]\ [(p-k-q)^2+m^2]\ k^2}\,, \tag 3.16
  $$ \vskip 10pt } \flushpar
where $\,\epsilon\,\equiv\,2-\omega\,$; $\ \epsilon\,$ occurs in the
integrand only because $\,\gamma^\mu \gamma_\mu = 2\omega\,$
(see step 2 above).

\medpagebreak
The \it zeros \rm and \it ones \rm at the right-hand sides of some
table entries are explained in the Key following Table 1.
The two entries marked ${\,}_{00}\,$ in Table 2 may be ignored, since it
can be demonstrated by the nested method (without the use of series)
that the corresponding integrals cancel each other exactly.
The three stars (*) in Tables 4 and 6 indicate integrals whose
contributions to the divergent parts of $I$ vanish due to symmetric
integration.  Hence only 48 integrals require evaluation.  All necessary
momentum-space integrals $J[P]$ are given in Appendix A,
leaving only the integrations over the parameters still to be done.

\baselineskip=19pt \settabs 14 \columns
\vskip 30pt
\centerline{\underbar{Table 1}:  Terms of $I$ with 5 denominator factors
including $\,q\!\cdot\!n\,$.}

\+&--------------------&&--------------------&&------------
&-----------&--------------------&&--------------------&&------------------\cr
\vskip -18pt \+&--------------------&&--------------------&&------------
&-----------&--------------------&&--------------------&&------------------\cr
\vskip -14pt \+&&& $\|$ && $|$ && $|$ && $|$ && $\|$ && $\,|$ \cr
\vskip -14pt \+& $ n\!\!\!/\,(p^2\!+\!m^2)$ &&&&&&
$\ \ \ 2+4\epsilon \,\ \ {\ \ }^0$ && $\ -2+2\epsilon \,\ \ {\ \ }^0$ &&
$\ \ \ \,4-4\epsilon \ \,\ {\ \ }^0$ \cr
\vskip -14pt \+&&& $\|$ && $|$ && $|$ && $|$ && $\|$ && $\,|$ \cr
\vskip -14pt \+&--------------------&&--------------------&&------------
&-----------&--------------------&&--------------------&&------------------\cr
\vskip -18pt \+&--------------------&&--------------------&&------------
&-----------&--------------------&&--------------------&&------------------\cr
\vskip -14pt \+&&& $\|$ && $|$ && $|$ && $|$ && $\|$ && $\,|$ \cr
\vskip -14pt \+& $\ \ \ m\, n\!\cdot\! k$ &&
$\ \ \ \,8-8\epsilon $   && $\ \ \ \ 4$ &&
$\ \ \ \ \ \ \ \ \ \ \ \,\ \ {\ \ }^0_0$ &&
$\ \ \ 8-8\epsilon \,\ \ {\ \ }^0_0$ && $\ \,-8+8\epsilon \ \,\ {\ \ }^0$ \cr
\vskip -14pt \+&&& $\|$ && $|$ && $|$ && $|$ && $\|$ && $\,|$ \cr
\vskip -14pt \+&--------------------&&--------------------&&------------
&-----------&--------------------&&--------------------&&------------------\cr
\vskip -14pt \+&&& $\|$ && $|$ && $|$ && $|$ && $\|$ && $\,|$ \cr
\vskip -14pt \+& $\ \ \ p\!\!\!/\ n\!\cdot\! k$ &&
$\ \ \ \,8-8\epsilon $   && $\ \ -4$ &&
$\ \ \ \ \ \ \ \ \ \ \ \,\ \ {\ \ }^0_0$ &&
$\ \ \ 8-8\epsilon \ \ \,{\ \ }^0_0$ && $\ \,-8+8\epsilon \ \,\ {\ \ }^0$ \cr
\vskip -14pt \+&&& $\|$ && $|$ && $|$ && $|$ && $\|$ && $\,|$ \cr
\vskip -14pt \+&--------------------&&--------------------&&------------
&-----------&--------------------&&--------------------&&------------------\cr
\vskip -14pt \+&&& $\|$ && $|$ && $|$ && $|$ && $\|$ && $\,|$ \cr
\vskip -14pt \+& $\ \ \ n\!\cdot\! p\ k\!\!\!/$ &&
$\ \,-4+4\epsilon $   && $\ \ 20-8\epsilon $ &&
$\ \ \ \ \ \ \ \ \ \ \ \,\ \ {\ \ }^0_0$ && $\ -4+4\epsilon \ \ {\ \ }^0_0$ &&
$\ \ \ \,4-4\epsilon \ \,\ {\ \ }^0$ \cr
\vskip -14pt \+&&& $\|$ && $|$ && $|$ && $|$ && $\|$ && $\,|$ \cr
\vskip -14pt \+&--------------------&&--------------------&&------------
&-----------&--------------------&&--------------------&&------------------\cr
\vskip -18pt \+&--------------------&&--------------------&&------------
&-----------&--------------------&&--------------------&&------------------\cr
\vskip -14pt \+&&& $\|$ && $|$ && $|$ && $|$ && $\|$ && $\,|$ \cr
\vskip -14pt \+& $\ \ \ n\!\cdot\! p\ q\!\!\!/$ &&
$\ \ \ \,4-4\epsilon \ \ {\ \ }_1$   && $\ \ -4\,\ \ \ \ \ \ \ {\ \ }_1$ &&
$\ \ \ 4+8\epsilon \,\ \ {\ \ }^0_0$ && $\ \ \ 4-4\epsilon \,\ \ {\ \ }^0_0$ &&
$\ \,-4+4\epsilon \,\ {\ \ }^0_{01}$ \cr
\vskip -14pt \+&&& $\|$ && $|$ && $|$ && $|$ && $\|$ && $\,|$ \cr
\vskip -14pt \+&--------------------&&--------------------&&------------
&-----------&--------------------&&--------------------&&------------------\cr
\vskip -14pt \+&&& $\|$ && $|$ && $|$ && $|$ && $\|$ && $\,|$ \cr
\vskip -14pt \+& $\ \ \ n\!\!\!/\ p\!\cdot\! q$ &&
$\ \ \ \ \ \ \ \ \ \ \,\ \ \ {\ \ }_1$ && $\ \ -4\,\ \ \ \ \ \ \ {\ \ }_1$ &&
$\ \ \ \ \ \ \ \ \ \ \ \,\ \ {\ \ }^0_0$ &&
$\ \ \ \ \ \ \ \ \ \ \ \,\ \ {\ \ }^0_0$ &&
$\ \ \ \ \ \ \ \ \ \ \ \ \ {\ \ }^0_{01}$ \cr
\vskip -14pt \+&&& $\|$ && $|$ && $|$ && $|$ && $\|$ && $\,|$ \cr
\vskip -14pt \+&--------------------&&--------------------&&------------
&-----------&--------------------&&--------------------&&------------------\cr
\vskip -18pt \+&--------------------&&--------------------&&------------
&-----------&--------------------&&--------------------&&------------------\cr
\vskip -14pt \+&&& $\|$ && $|$ && $|$ && $|$ && $\|$ && $\,|$ \cr
\vskip -14pt \+& $\ \ \ n\!\!\!/\ k\!\cdot\! q$ &&
$\ \ \,\ \ \ \ \ \ \ \ \ \ \ {\ \ }_1$ &&
$\,\ \ \ \ \ \ \ \ \ \ \ \ \ {\ \ }_1$ &&
$\,\ \ \ \ \ \ \ \ \ \ \ \ \ {\ \ }^0_0$ &&
$\,\ \ \ \ \ \ \ \ \ \ \ \ \ {\ \ }^0_0$ &&
$\ \,-4+4\epsilon \,\ {\ \ }^0_{01}$ \cr
\vskip -14pt \+&&& $\|$ && $|$ && $|$ && $|$ && $\|$ && $\,|$ \cr
\vskip -14pt \+&--------------------&&--------------------&&------------
&-----------&--------------------&&--------------------&&------------------\cr
\vskip -14pt \+&&& $\|$ && $|$ && $|$ && $|$ && $\|$ && $\,|$ \cr
\vskip -14pt \+& $\ \ \ n\!\cdot\! k\ q\!\!\!/$ &&
$\ \ \ \,\ \ \ \ 4\epsilon \ \ \ {\ \ }_1$ &&
$\,\ \ \ \ \ \ \ \ \ \ \ \ \ {\ \ }_1$ &&
$\ \ \ 4-4\epsilon \,\ \ {\ \ }^0_0$ && $\ -8+8\epsilon \ \ {\ \ }^0_0$ &&
$\ \,\ \ 4-4\epsilon \,\ {\ \ }^0_{01}$ \cr
\vskip -14pt \+&&& $\|$ && $|$ && $|$ && $|$ && $\|$ && $\,|$ \cr
\vskip -14pt \+&--------------------&&--------------------&&------------
&-----------&--------------------&&--------------------&&------------------\cr
\vskip -14pt \+&&& $\|$ && $|$ && $|$ && $|$ && $\|$ && $\,|$ \cr
\vskip -14pt \+& $\ \ \ n\!\cdot\! k\ k\!\!\!/$ &&
$\ \,-4+4\epsilon $   && $\ -4+4\epsilon \ \ {\ \ }_0$ &&
$\ \ \ \ \ \ \ \ \ \ \ \ {\ \ }^0_{00}$ &&
$\ \ \ \ \ \ \ \ \ \ \ \ {\ \ }^0_{00}$ &&
$\ \ \ \ \ \ \ \ \ \ \ \ \ {\ \ }^0_{00}$ \cr
\vskip -14pt \+&&& $\|$ && $|$ && $|$ && $|$ && $\|$ && $\,|$ \cr
\vskip -14pt \+&--------------------&&--------------------&&------------
&-----------&--------------------&&--------------------&&------------------\cr
\vskip -18pt \+&--------------------&&--------------------&&------------
&-----------&--------------------&&--------------------&&------------------\cr
\vskip -14pt \+&&& $\|$ && $|$ && $|$ && $|$ && $\|$ && $\,|$ \cr
\vskip -14pt \+&&& $\|$ \ nq ${}_\wedge $Kk && $|$ \ n Q${}_\wedge $Kk && $|$
 \ nqQ${}_\wedge $K && $|$ \ nqQ${}_\wedge $ k && $\|$ \ nqQ Kk && $\,|$ \cr

\newpage
\flushpar \underbar{Key to a typical entry}:

\hskip 67pt -- value of $\,a\,$ at which $\,\int J_1\,$ diverges.
 (See Section 4.)
\vskip -14pt \hskip 1.5pt ---------------\ \ \ \,/
\vskip -14pt $|$ \hskip 44pt $|\!\leftarrow $
\vskip -14pt \ \ $ 4-4\epsilon \,{\ \ }^0_{01} $
\vskip -14pt $|$ \hskip 44pt $|\!\leftarrow\! $--- values of $\,a\,$
at which $\,\int J_0\,$ diverges.  ${\ \ }_{00}\,$ indicates that
\vskip -14pt \hskip 1.5pt ---------------
\vskip -10pt \hskip 2pt \ \ \ \ $\uparrow $
\vskip -14pt \hskip 90pt the degree is --1 at $\,a=0$.  (See Section 4(b).)
\vskip -14pt \ \ \ \ \ $|$
\vskip  -9pt \ \ \ \ \ $|$
\vskip -14pt \ \ \ \ \ \,--- coefficient of the term
\ $(\,\epsilon\,\equiv\,2-\omega\,)$.
\vskip 38pt

\settabs 15 \columns
\centerline{\underbar{Table 2}:  Terms of $I$ with 4 denominator factors
including $\,q\!\cdot\!n\,$.}

\+&-----------------&&-----------------&&-----------------
&&-----------------&&-----------------&&-----------------&&---------\cr
\vskip -14pt \+&& $\|$ && $|$ && $|$ && $|$ && $\|$ && $\ |$ && $|$ \cr
\vskip -14pt \+& \ \ $ n\!\!\!/$ & \ \ \ \ \ \ \ $2\epsilon\,
 {\ \ \ \ }_1$ && \ \ \ \ $-2\epsilon {\ \ \ \ }_{01}$ &&
 \ \ --\,2\,--\,2$\epsilon \,{\ \ \ }_0$ &&
 \ \ --\,4\,--\,2$\epsilon $ && \ \ --\,2\,+2$\epsilon {\ \ \ }_{00}$ &&
 \,\ \ 2\,--\,2$\epsilon \,{\ \ \ }_{00}$ \cr
\vskip -14pt \+&& $\|$ && $|$ && $|$ && $|$ && $\|$ && $\ |$ && $|$ \cr
\vskip -14pt \+&-----------------&&-----------------&&-----------------
&&-----------------&&-----------------&&-----------------&&---------\cr
\vskip -14pt \+&& $\|$ && $|$ && $|$ && $|$ && $\|$ && $\ |$ && $|$ \cr
\vskip -14pt \+&& $\|$ \ n \ ${}_\wedge $Kk && $|$ \ n Q Kk && $|$
 \ n Q${}_\wedge $ k && $|$ \ nq ${}_\wedge $K && $\|$ \ nqQ${}_\wedge $
 && $\ |$ \ nqQ K && $|$ \cr
\vskip 30pt

\settabs 13 \columns
\centerline{\underbar{Table 3}:  Covariant-gauge terms of $I$.}

\+&---------------------&&---------------------&&---------------------
&&---------------------&&---------------------&&------------\cr
\vskip -18pt
\+&---------------------&&---------------------&&---------------------
&&---------------------&&---------------------&&------------\cr
\vskip -14pt \+&& $\|$ && $|$ && $\!\!\!\!|$ && $|$ && $\|$ && $\,\ \ |$ \cr
\vskip -14pt \+&\ \ \ m & \ \ 2\,--2$\epsilon$\,+2$\epsilon^2 {\ \ }_1$ &&
\ \ \ \ \ \ \ \ \ \ \ \ \ ${\ \ }_1$ &&
\ \,8\ \ \ \ \ --\,8$\epsilon^2 {\ \ \ }_0$ &&
\ \,2\,--2$\epsilon$\,+2$\epsilon^2 {\,\ \ }_0$ &&
\ \ --2\,+2$\epsilon$\,--2$\epsilon^2 {\ \ }_{01}$ \cr
\vskip -14pt \+&& $\|$ && $|$ && $\!\!\!\!|$ && $|$ && $\|$ && $\,\ \ |$ \cr
\vskip -14pt
\+&---------------------&&---------------------&&---------------------
&&---------------------&&---------------------&&------------\cr
\vskip -14pt \+&& $\|$ && $|$ && $\!\!\!\!|$ && $|$ && $\|$ && $\,\ \ |$ \cr
\vskip -14pt \+&\ \ \ $ p\!\!\!/$ & \ \ 4\,--2$\epsilon$\,+2$\epsilon^2
 {\ \ }_1$ && \ \ \ \ \ \ \ \ \ \ \ \ \ ${\ \ }_1$ &&
\ \,8\,--\,8$\epsilon$\,--\,8$\epsilon^2 {\,\ \ }_0$ &&
\ \,4\,--2$\epsilon$\,+2$\epsilon^2 {\,\ \ }_0$ &&
\ \ --4\,+2$\epsilon$\,--2$\epsilon^2 {\ \ }_{01}$ \cr
\vskip -14pt \+&& $\|$ && $|$ && $\!\!\!\!|$ && $|$ && $\|$ && $\,\ \ |$ \cr
\vskip -14pt
\+&---------------------&&---------------------&&---------------------
&&---------------------&&---------------------&&------------\cr
\vskip -14pt \+&& $\|$ && $|$ && $\!\!\!\!|$ && $|$ && $\|$ && $\,\ \ |$ \cr
\vskip -14pt \+&\ \ \ $ q\!\!\!/$ &
 \ \ 4\,--\,4$\epsilon \ \ \ \ \ \ {\ \ }_1$ &&  \ \,--\,8\,+\,8$\epsilon
 \ \ {\ \ }_1$ && \!--\,8+4$\epsilon$\,+4$\epsilon^2 {\ \ }_0$ &&
\ \,--\,8\ \ \ +4$\epsilon^2 {\,\ \ }_0$ &&
\ \ \ \,4\,--\,4$\epsilon\ \ \ \ \ \ {\ \ }_{01}$ \cr
\vskip -14pt \+&& $\|$ && $|$ && $\!\!\!\!|$ && $|$ && $\|$ && $\,\ \ |$ \cr
\vskip -14pt
\+&---------------------&&---------------------&&---------------------
&&---------------------&&---------------------&&------------\cr
\vskip -18pt
\+&---------------------&&---------------------&&---------------------
&&---------------------&&---------------------&&------------\cr
\vskip -14pt \+&& $\|$ && $|$ && $\!\!\!\!|$ && $|$ && $\|$ && $\,\ \ |$ \cr
\vskip -14pt \+&& $\|$ \ \ q ${}_\wedge $Kk && $|$ \ \ \ Q${}_\wedge $Kk
 && $\!\!\!\!|$ \ \ \ \ qQ${}_\wedge $K && $|$ \ \ qQ${}_\wedge $ k
 && $\|$ \ \ qQ Kk && $\,\ \ |$ \cr
\vskip 30pt

\settabs 13 \columns
\centerline{\underbar{Table 4}:  Terms of $I$ with 5 denominator factors
including $\,q\!\cdot\!n\,$ and $\,k\!\cdot\!n\,$.}

\+&---------------------&&---------------------
&&---------------------&&---------------------&&-----------\cr
\vskip -14pt \+&& $\|$ && $|$ && $\|$ && $|$ && $\|$ \cr
\vskip -14pt \+& \ $ n\!\!\!/\ n\!\cdot\! p$ &
 \ \ \ \ \ 4 \ \ \ \ \ \ ${\ \ }^0$ && \ \ \ $-4 \ \ \ \ \ \ {\ \ }^0$ &&
 \ \ \ \ \ 8 \ \ \ \ \ \ \ * && \ \ \ $-4$ \ \ \ \ \ \ \ * \cr
\vskip -14pt \+&& $\|$ && $|$ && $\|$ && $|$ && $\|$ \cr
\vskip -14pt \+&---------------------&&---------------------
&&---------------------&&---------------------&&-----------\cr
\vskip -14pt \+&& $\|$ && $|$ && $\|$ && $|$ && $\|$ \cr
\vskip -14pt \+&& $\|$ \ nqQ K n && $|$ \ nqQ${}_\wedge $ \ n && $\|$
 \ n Q${}_\wedge $K n && $|$ \ nq ${}_\wedge $K n && $\|$ \cr

\vskip 30pt
\hskip 48pt \underbar{Table 5} \hskip 132pt \underbar{Table 6}

\+&---------------------&&-----------------------&&
&&---------------------&&--------------------\cr
\vskip -14pt \+&&& $\ \ \ \ \ |$ && $\ \ \ \ |$ && && $|$ && $|$ \cr
\vskip -14pt \+& \ $ n\!\cdot\! k\ k\!\!\!/\,(p^2+m^2)$
 && \ \ \ \ \ \ \ \ \ $4-4\epsilon \ \ \ {\ }^0$ && &&
 $ n\!\!\!/\ n\!\cdot\! p\ (q\!\cdot\! k)^2$
 && \ \ \ \ 4 \ \ \ \ \ \ \ * \cr
\vskip -14pt \+&&& $\ \ \ \ \ |$ && $\ \ \ \ |$ && && $|$ && $|$ \cr
\vskip -14pt \+&---------------------&&-----------------------&&
&&---------------------&&--------------------\cr
\vskip -14pt \+&&& $\ \ \ \ \ |$ && $\ \ \ \ |$ && && $|$ && $|$ \cr
\vskip -14pt \+&&& $\ \ \ \ \ |$ \ nqQ${}_\wedge $Kk &&
 $\ \ \ \ |$ && && $|$ \ nqQ${}_\wedge $Kkn && $|$ \cr

\baselineskip=20pt
\leftline{\bf 4. Finding the Divergences }
\leftline {\bf \ \ (a) Integration over infinite-parameter space }
\vskip 15pt
   In order to complete the parameter change from $\{\alpha_j\}$ to
$\{A;\lambda,\beta,G,b,h,a\}$, we have to apply the transformation (3.4)
to the integrations on the right-hand side of Eq\. (3.1) for each
of our 48 integrals.  Integrals containing
all seven ``factors'' n, q, Q, ${}_\wedge$, K, k, n \,in their denominators
then transform as follows:
  \flushpar {\hsize=371pt \vskip -12pt $$ \align
  I_E[f]\ &=\ \int_0^\infty d\alpha_1\,\dots\int_0^\infty d\alpha_7\,J[P]
  \\ \vspace{12pt}
  &=\ \int_0^\frac12\! dG \int_G^{1-G}\! da \int_G^{1-{\dsize a}}\! d\lambda
  \int_G^{\dsize a}\! dh \int_G^{\dsize\lambda}\! d\beta \int_G^{\dsize h}\! db
  \int_0^\infty\! A^6\,dA\,J[P]\,.\quad \tag 4.1 \endalign
  $$ \vskip 8pt } \flushpar
But let's suppose now that a certain denominator factor is missing from the
simplified integrand $\,f\,$ in Eq\. (3.1).  In that case, as mentioned in
Section 3(c), we must set the corresponding $\,\alpha\,$ parameter to zero in
Eqs\. (3.4), and omit integration over this parameter in Eq\. (3.1).  To see
the effect of these changes on the \it transformed \rm integral (4.1),
consider the example (3.16), in which $\,f= q\!\!\!//$(nqQ${}_\wedge$k).
Since the fifth and seventh denominator factors are absent in this case,
we put $\,\alpha_5=\alpha_7=0\,$ to get
  \flushpar {\hsize=371pt \vskip -10pt $$ \align
  I_E[f]\ &= \int_E d^{2\omega}q \int_E d^{2\omega}k \frac{q\!\!\!/}
                 {\text{nqQ}{}_\wedge\text{ k}}
  \\ \vspace{12pt}
      &= \int_0^\infty\! d\alpha_1\int_0^\infty\! d\alpha_2\int_0^\infty\!
         d\alpha_3\int_0^\infty\! d\alpha_4\ \int_0^\infty\! d\alpha_6\,
         J\left[\frac{-n^*\!\!\cdot\!q\ q\!\!\!/}
                        {n_0^2}\right]_{\dsize\alpha_5=\alpha_7=0}
  \\ \vspace{12pt}
    &= \int_0^\frac12\! dG \int_G^{1-G}\! da \int_G^{1-{\dsize a}}\! d\lambda
       \int_G^{\dsize\lambda}\! d\beta\ \biggl[\,J_A(\lambda,\beta,G,b,h,a)
            \,\biggr]_{\dsize b\to G\atop\dsize h\to a}\,, \tag 4.2
  \\ \vspace{20pt}
    \text{with} &\qquad J_A \,\equiv \int_0^\infty\!\! A^4\,dA\,
    J\left[\frac{-n^*\!\!\cdot\! q\ q\!\!\!/}{n_0^2}\right]
    \ = \ -\int_0^\infty\!\! dA\,\frac{A^4}{n_0^2}\,\gamma^\mu\,
    J\bigl[ n^*\!\!\cdot\! q\ q_\mu\bigr]\,. \tag 4.3 \endalign
  $$ \vskip 6pt }

\newpage \flushpar
In the next few pages, we shall concentrate on the detailed evaluation of the
integral (4.2).

\medpagebreak
In view of Eqs\. (A.4) and (A.8), we find that $J_A$ in Eq\. (4.3) becomes
  \flushpar {\hsize=371pt \vskip -8pt $$
  J_A \ = \ -\,\frac{\pi^{4-2\dsize\epsilon}}{n_0^2} \int_0^\infty\!\! dA\,
          \frac{A^{2\dsize\epsilon}\ \bold e^{-AH}}
          {D_\parallel D_\perp^{1-\dsize\epsilon}}
          \left( n^*\!\!\cdot\! s\ s\!\!\!/\,+
          \,\frac{a n\!\!\!/^*}{2AD_\parallel}\right). \tag 4.4
  $$ \vskip 8pt } \flushpar
Integration over $A$ is straightforward.  Note that
Eq\. (4.4) shows the entire A-dependence of the integrand, since
$\,r,\ s,\ D_\parallel,\ D_\perp,\,$ and $H$ have been defined
so as to be independent of $A$.  Using Eqs\. (3.14), together with
  \flushpar {\hsize=371pt \vskip -8pt $$
  \int_0^\infty t^x\,\bold e^{-t}\,dt \ = \ \Gamma(x+1)\,,
  \qquad\text{Re}(x+1)>0\,, \tag 4.5
  $$ \vskip 5pt } \flushpar
we find that
  \flushpar {\hsize=371pt \vskip -16pt $$
  \biggl[\,J_A\,\biggr]_{\dsize b\to G\atop\dsize h\to a}\ =
  \ -\,\frac{\pi^{4-2\dsize\epsilon}}{n_0^2}\,\bigl[\,\Gamma(2\epsilon)\ J_0
  \ +\ \Gamma(1+2\epsilon)\ J_1\ +\ \Gamma(2+2\epsilon)\ J_2\,\bigr]\,,\tag 4.6
  $$ \vskip 5pt} \flushpar
where
  \flushpar {\hsize=371pt \vskip -16pt $$
  \qquad\quad J_0 = \ \frac{ n\!\!\!/^* aH^{-2\dsize\epsilon}}
    {2 D_\parallel^2 D_\perp^{1-\dsize\epsilon}}\,,
  \qquad J_1 = \ \frac{ n^*\!\!\cdot\! p \,(a\beta-G^2)^2}
  {D_\parallel^2 D_\perp^{1-\dsize\epsilon}H} \left(\frac
  { p\!\!\!/_\parallel}{D_\parallel}+\frac{ p\!\!\!/_\perp}{D_\perp}
  \right)H^{-2\dsize\epsilon}\,,  \tag 4.7
  $$ \vskip 10pt } \flushpar
and $J_2=0$.  Similar results may be obtained for the other integrals
in Tables 1 to 6; only the expressions for $J_0,\,J_1$, and $J_2$ vary.

\medpagebreak
As explained in the \underbar{Introduction}, the first pole term
$\,\Gamma(2\epsilon)\,J_0\,$ in Eq\. (4.6) arises from integration over the
type I parameter $A$.  This divergence may be traced back to the fact that
the original integral $I$ is UV-divergent
with respect to $\,q\,$ and $\,k\,$ taken together.  Additional UV \it
subdivergences \rm will emerge when the $J_0$ and $J_1$ terms are integrated
over the \it finite parameters \rm $\,\lambda,\beta,G,b,h,\,$ and $\,a\,$
(called type II parameters in the \underbar{Introduction}).  These
subdivergences arise from the two loops taken separately.


\medpagebreak
In view of the appearance of $D_\parallel,\,D_\perp,\,$ and $H$ in Eq\.
(4.7), exact integration of $J_0$ and $J_1$ over the finite parameters
would pose a major challenge.  Fortunately, however, we are only
interested in contributions to $I$ which diverge as $\,\omega\to 2$;
i.e., in the terms with negative indices in the Laurent series
  \flushpar {\hsize=371pt \vskip -12pt $$
  I(\epsilon)\ =\ I_{-t}(p,n;m)\epsilon^{-t}\,+
                \ I_{1-t}(p,n;m)\epsilon^{1-t}\,+\,\dots\,, \tag 4.8
  $$ \vskip 3pt } \flushpar
where $\,t\,$ is an integer which, in our case, happens to be 2.
The divergences occur only at certain boundaries of the
region of integration in parameter space.  We can extract the divergent
terms exactly by integrating the first one or two terms of suitably chosen
series which converge to $\,J[P]\,$ near these boundaries.  In choosing
these series, we must be careful to avoid the types of problems which
plague the nested method.

\medpagebreak
In the case of the $J_0$ term, one boundary at which the parameter
integrations diverge as $\,\omega\to 2\,$
is just $\,A=0$; this fact explains why integration over $A$ generates
the divergent factor $\,\Gamma(2\epsilon)\,$ in this term.  Of course, the
presence of this factor implies that the coefficient $I_{-1}\,$ in Eq\.
(4.8) will depend on the \it finite \rm part of the integral of $J_0$ over
the remaining parameters.  It would seem, therefore, that the $J_0$ terms
will have to be integrated over the \it entire \rm region of integration
$\Phi$ of these parameters.

\medpagebreak
Fortunately, things are not quite as bad as they seem, since the most
complicated part of the $J_0$ term is the factor $ H^{-2\dsize\epsilon} $.
If we use the expansion
  \flushpar {\hsize=371pt \vskip -15pt $$
  H^{-2\dsize\epsilon}\, = \ \bold e^{-2\dsize\epsilon\ln H}\, = \ \ 1\,-\,
  2\epsilon\ln H\,+\,2\epsilon^2 (\ln H)^2\,-\ \dots\ ,\tag 4.9
  $$ \vskip 3pt } \flushpar
it would appear that we only need to keep the first term of the series
when integrating over the whole region $\Phi$, since the common factor of
$\,2\epsilon\,$ in the other terms will cancel the divergence arising
from the factor $\,\Gamma(2\epsilon)$ in Eq\. (4.6).  However, it was
\pagebreak
just this kind of reasoning that led to the catastrophe in the nested
method exemplified by Eq\. (2.11).  To avoid a similar catastrophe here,
we must verify that $\int_\Phi J_0$ is convergent at \it all zeros and poles
\rm of $H$ before we may use the series (4.9).  It follows from definition
(A.9) that $H$ has no poles in $\Phi$, but $H$ does go to \it zero \rm
when $b+\beta-G\to 0$.  In order to determine whether the integral of any
$J_0$ term diverges at any of these zeros as $\,\omega\to 2$, and in order
to enable us to find all divergent contributions to the final result $I$,
we must systematically determine the location in finite-parameter space
of every subdivergence of every $J_0$, $J_1$, and $J_2$ integral.
This task will be tackled in the next subsection.

\vskip 30pt
\leftline {\bf \ \ (b) Subdivergences }
\vskip 15pt

Divergences in the finite-parameter integrations, as $\,\omega\to 2$,
could potentially occur at the zeros of any of the three factors
$D_\parallel,\ D_\perp,$ and $H$ which appear in the denominators of the
$J_0,\ J_1,$ and $J_2\,$ terms.  These zeros occur on certain boundaries
of the integration region $\Phi$, such as $\,a=0,\ \lambda=h=G$, etc.
Specifically,
  \flushpar {\hsize=371pt \vskip -12pt $$
  \qquad\quad \left. \matrix
  D_\parallel \to 0\quad\text{\ \ \ linearly\ \ \ \ \ \ \ \,\ \ if}\quad
  a\to 0,\ \ a\to 1,\ \ \text{or}\ \ \,a,G\to \tfrac12\,, \\ \vspace{8pt}
  D_\perp \to 0\ \left\{ \matrix \text{\ quadratically\ \,\ \ if\ \ \ \it both}
                   \ \ \ \lambda,\ h\to 0\,,\ \ \qquad\qquad\qquad
  \\ \vspace{8pt} \text{or else linearly\ \ if}\quad \lambda\to 0,\ \ h\to
    0,\ \ \text{or}\ \ \,\lambda,h\to G\,, \endmatrix \right.
  \endmatrix \right\} \tag 4.10a
  $$ \vskip 12pt } \flushpar
(cf\. Eqs\. (3.7) and (4.1)).
To locate the zeros of $H$, we observe that the exponent of the integrand in
Eq\. (3.3) is less than or equal to zero by construction, and equal to
$\,-AH\,$ when $\,\bold y=0\,$ (cf\. Eqs\. (A.9) and (3.9)).
Hence $\,H\geq 0\,$ even if the mass $\,m\,$ vanishes, while for $\,m\neq 0,$
  \flushpar {\hsize=371pt \vskip -12pt $$
  H\to 0\quad\text{linearly,\ \ \ if and only if\ \ both}\quad b,\beta\to 0\,.
  \tag 4.10b
  $$ \vskip 5pt }

Depending on the specific form of the integrand, a given
finite-parameter integral could conceivably diverge at the intersection
of some particular subset of the above boundaries, but not at any other
points.  Accordingly, we must consider all possible intersections of these
boundaries for every term of every finite-parameter integrand.  The set
of possibilities which needs to be examined is severely
constrained by two chains of inequalities, which follow from Eqs\. (3.4):
  \flushpar {\hsize=371pt \vskip -18pt $$
  a\ \geq\ h\ \geq\ b\ \geq\ G\ \geq 0\qquad\text{and}\qquad 1-a\ \geq
  \ \lambda\ \geq\ \beta\ \geq\ G\ \geq 0\,. \tag 4.11
  $$ \vskip 2pt } \flushpar
If any particular parameter in either of these chains goes to zero,
then all parameters to its right in the chain must also approach zero.

\medpagebreak
In order to determine whether one of our integrals is divergent at a
particular boundary of $\Phi$, we observe that the integral of a rational
function of the parameters diverges at some point in finite-parameter
space \it only if the degree of the integrand \rm (including the measure
$\,dG\,da\dots\,$) \it is less than or equal to zero at this point. \rm
The degree of the integrand may be calculated
according to the following rules: \vskip 8pt
\item{1) } A linear function of a parameter has degree 1 if it goes to
zero at the point in question; otherwise the function has degree 0.
\item{2) } The degree of a product (quotient) is the sum (difference) of
the degrees of its parts.
\item{3) } The degree of a sum or difference, in the numerator, is the
\it minimum \rm of the degrees of its parts.  For example, if $\,f\to 0$
linearly and $\,g\to 0$ quadratically, then $\,f+g\to 0$ linearly.  (If the
parts have equal degree, this rule may predict divergences which eventually
cancel each other.)
\item{4) } The degrees of the denominator factors $D_\parallel,\,D_\perp,$
and $H$ are given by Eqs\. (4.10) for all boundaries except those at which the
degrees are zero.
\item{5) } The degree of the measure is the dimensionality of
finite-parameter space minus the dimensionality of the boundary in question
(not counting dimensions associated with absent parameters, such as $\,b\,$
and $\,h\,$ in Example (4.2)).  For instance, at the boundary $\,a=0\,$,
inequalities (4.11) imply that $G$ must also be zero; hence the degree of the
measure $\,dG\,da\,d\lambda\,d\beta\,$ at this boundary is $\,4-2=2\,$.  To see
why, consider a transformation of $\,a\,$ and $\,G\,$ to polar co-ordinates.

\bigpagebreak
As an example, consider the three terms comprising $J_0$ and $J_1$ in
Eqs\. (4.7).  If we multiply these
terms by $\,dG\,da\,d\lambda\,d\beta\,$ to complete the integrands, we
find that the degrees of these integrands are as shown in Tables 7 and 8.
Each number in these tables represents the degree of the integrand
at the boundary at which the parameters, shown above it and to the left,
go to zero.  From these tables we see that all three terms in Eqs\. (4.7)
have subdivergences at the boundary where
$\,a\to 0\,$ but \it not \rm $\,\beta\to 0\,$ or $\,\lambda\to 0$.

\vskip 20pt
\item{} \underbar{Table 7}:  Degree of $J_0 \qquad\qquad\qquad$
        \underbar{Table 8}:  Degrees of $J_1$ terms \vskip -5pt
\item{} at various boundaries.
$\quad\qquad\qquad\qquad\qquad$ at various boundaries.
\vskip 11pt \flushpar \ ------------------------------
$\ \ \ \ \ $ ------------------------------
$\ \ \ \ \ $ ------------------------------ \vskip -49pt
$$\matrix  &\!\!|\!\!& G &\beta &\lambda & 1-a &\!\!|\!\!&&&\!\!|\!\!& G &
\beta &\lambda & 1-a &\!\!|\!\!&&&\!\!|\!\!& G &\beta &\lambda & 1-a &\!\!|
\\ \vspace{-4pt}
&\!\!|\!\!&&&&&\!\!|\!\!&&&\!\!|\!\!&&&&&\!\!|\!\!&&&\!\!|\!\!&&&&&\!\!|
\\ \vspace{-4pt}
G &\!\!|\!\!& 1 & 2 & 2 & 1 &\!\!|\!\!&& G &\!\!|\!\!& 1 & 3 & 3 & 1
                            &\!\!|\!\!&& G &\!\!|\!\!& 1 & 3 & 2 & 1 &\!\!|
\\ a &\!\!|\!\!& 0 & 1 & 1 &&\!\!|\!\!&& a &\!\!|\!\!& 0 & 2 & 2 &
                            &\!\!|\!\!&& a &\!\!|\!\!& 0 & 2 & 1 &   &\!\!|
\endmatrix $$ \vskip 20pt

\medpagebreak
When tables, similar to Tables 7 and 8, are constructed for
every term of every finite-parameter integral which arises in the
evaluation of $I$, it is found that subdivergences
occur only at the following two boundaries of $\Phi$: \vskip 8pt
\item{1) } $a\to 0\,$ and \it not \rm $\,\beta\to 0$\ \ (corresponding to
divergent $\,k\,$ integration).
\item{2) } $a\to 1\,$ and \it not \rm $\,b\to 0$\ \ (corresponding to
divergent $\,q\,$ integration). \vskip 10pt \flushpar
Each integral which diverges at one or both of these boundaries is
marked in Tables 1 to 6 with ${\,}^0\,$ and/or ${\,}^1$,
as explained in the Key.  Note that \it neither \rm of these types of
divergence occurs at a boundary where $\,b\,$ and $\,\beta\,$
\it both \rm go to zero.  In particular, we see from Eq\. (4.10b) that
there are no subdivergences at zeros of $H$.  Therefore, it is safe
to use the series (4.9) in all cases, as we shall do from now on.

\medpagebreak
To calculate the divergent contributions to $I$ which arise at either of
the boundaries $\,a=0\,$ or $\,a=1\,$, we expand each integrand
$J_i\ (i=0,1,2)$ as a series in powers of the parameters which go to zero at
that boundary, and keep only those terms whose integrals diverge.
(We justify this procedure by using the five rules given above to show that
the integral of the difference between $J_i$ and the terms we keep
is convergent at all boundaries in every case.)
In all but two of the cases, the degree of the original
integrand is zero at the subdivergence, so that only the leading term of the
series is needed.  For each of the other two cases (marked ${\,}_{00}\,$
in Table 2), the degree of the $J_0$ integrand is $-1$, and
two terms would be needed.  Fortunately, we are saved from
having to integrate these terms, since these two exceptional
integrals cancel each other exactly (cf\. Section 3(c)).

\vskip 30pt
\leftline{\bf 5. Finite-Parameter Integration }
\leftline {\bf \ \ (a) Integrations at subdivergences }
\vskip 15pt
Employing the two series expansions discussed in Section 4,
we can now complete the finite-parameter integrations to the extent necessary
to obtain exact expressions for the coefficients $I_{-1}\,$ and $I_{-2}\,$
in Eq\. (4.8).  Proceeding with Example (4.2), we first substitute the series
for $H^{-2\dsize\epsilon}$ in Eq\. (4.9) into Eqs\. (4.7).  We then
use Eq\. (4.6) and the identity $\,\Gamma(z+1)=z\Gamma(z)$ to obtain
\vskip -6pt $$
  \biggl[\,J_A\,\biggr]_{\dsize b\to G\atop\dsize h\to a} \ = \ -\,
    \frac{\pi^{4-2\dsize\epsilon}}{n_0^2}\,\bigl[\,\Gamma(2\epsilon)\ Y_0\ +
    \ \Gamma(1+2\epsilon)\ Y_1\ +\ \Gamma(2+2\epsilon)\ Y_2\,\bigr]\ +
    O(\epsilon),
$$ \flushpar {\hsize=371pt \vskip -25pt $$ \tag 5.1 $$
$$ \align &\ \\ \vspace{-15pt} \text{where} \qquad
  Y_0 \ &= \ \frac{ n\!\!\!/^*\,a}
    {2D_\parallel^2 D_\perp^{1-\dsize\epsilon}}\,, \tag 5.2a
  \\ \vspace{10pt}
  Y_1 \ &= \ \frac{ n^*\!\!\cdot\! p
    \,(a\beta-G^2)^2}{D_\parallel^2 D_\perp^{1-\dsize\epsilon}H} \left(\frac
    { p\!\!\!/_\parallel}{D_\parallel}\,+\,\frac{ p\!\!\!/_\perp}{D_\perp}
    \right)\,-\,\frac{ n\!\!\!/^*\,a\,\ln H}
    {2D_\parallel^2 D_\perp^{1-\dsize\epsilon}}\,, \tag 5.2b \endalign
  $$ \vskip 10pt } \flushpar
and $Y_2=0$.  Similar results may be obtained for the other integrals
in Tables 1 to 6; only the expressions for $Y_0,\,Y_1$, and $Y_2$ vary.
It remains to integrate these expressions over the finite parameters.
Since $Y_1$ and $Y_2$ are not multiplied by divergent
$\Gamma$-functions, they contribute to $I_{-1}$ only through the
subdivergences in their finite-parameter integrals.  The single $Y_2$
integral has \it no \rm subdivergences, and may be ignored.
The $O(\epsilon)$ term in Eq\. (5.1) may also be ignored, since there are
no subdivergences at the zeros of $H$, as explained in Section 4(b).

\medpagebreak
Returning to our example in Eqs\. (5.2), we now compute the divergent part of
the integral of $Y_1$ over the finite parameters $\,\beta,\,\lambda,\,a\,$
and $G$ (cf\. Eq\. (4.2)).  We recall from Section 4(b) that all
subdivergences of Eq\. (4.2) occur at the boundary $\,a=G=0$.  Therefore,
we expand $Y_1$ as
a series in the parameters $\,a\,$ and $\,G$, and integrate only the divergent
leading term.  To derive this term, we simply replace $H$ by its leading term
$H_0\equiv H_{{\dsize a}\to 0}\,$, and drop $G^2$ and
$(1-a)$ from $D_\parallel,\,D_\perp$, and from the numerator in Eq\. (5.2b).
In this way we obtain
  \flushpar {\hsize=371pt \vskip -18pt $$ \align
  Y_1\, &=\, a^{{\dsize\epsilon}-2}\,E\ +O(a^{{\dsize\epsilon}-1}), \tag 5.3
  \\ \vspace{10pt} \text{where}\qquad
  E\,&\equiv\,\frac{ n^*\!\!\cdot\! p\,\beta^2} {\lambda^{1-\dsize\epsilon}H_0}
    \left( p\!\!\!/_\parallel+\,\frac{ p\!\!\!/_\perp}\lambda\right)\,-
    \,\frac{ n\!\!\!/^*\ln H_0} {2\,\lambda^{1-\dsize\epsilon}} \,. \tag 5.4
  \endalign
  $$ \vskip 8pt } \flushpar
Applying Eqs\. (A.9), (3.14), and (3.7), and recalling that $\,a=h\,$ in
our example, and that $\ 0\leq b\leq a\,$ (cf\. inequality (4.11)), we find
  \flushpar {\hsize=371pt \vskip -10pt $$
  \qquad\qquad H_0 \ = \  p_\parallel^2\,\beta
    \left(R-\beta-\,\frac{\beta S}{\lambda} \right),\quad \ \ \ \left.
  \matrix \text{where}\ & R\equiv\,( p^2+m^2)/ p_\parallel^2\,,
  \\ \vspace{8pt}
  \text{and} & S\equiv\, p_\perp^2/ p_\parallel^2\,.
  \endmatrix \ \ \ \right\} \!\!\! \tag 5.5
  $$ \vskip -10pt } \flushpar
Integration of Eq\. (5.3) over $\,\beta,\,\lambda,\,a$,
and $\,G\,$ in accordance with Eq\. (4.2) yields
  \flushpar {\hsize=371pt \vskip -20pt $$ \align
  \int_\Phi Y_1\ &= \int_0^{\frac12}\! dG \int_G^{1-G}\! da
     \int_G^{1-{\dsize a}}\!\!\! d\lambda \!\int_G^{\dsize\lambda}\!\!
     d\beta \ a^{{\dsize\epsilon}-2}\,E \quad +\ \text{finite terms}\,,
  \\ \vspace{8pt}
  &= \int_0^{\frac12}\! dG \int_G^{1-G} a^{{\dsize\epsilon}-2}\, da
     \ (L_0-L_1) \quad +\ \text{finite terms}\,, \tag 5.6
  \\ \vspace{8pt}  \text{where}\qquad
  L_0\ &\equiv\ \int_0^1 d\lambda \int_0^{\dsize\lambda} d\beta\ E\,,\tag 5.7a
  \\ \vspace{8pt}
  L_1\ &\equiv\ \int_{1-{\dsize a}}^1 d\lambda \int_G^{\dsize\lambda} d\beta
     \ E\ +\int_0^G d\beta \int_{\dsize\beta}^1 d\lambda \ E\,, \tag 5.7b
  \endalign
  $$ \vskip 6pt } \flushpar
and $E$ is defined by Eq\. (5.4).
Integral (5.7a) is convergent even for $\,\omega=2\ (\epsilon=0)$,
and may be evaluated in the $\,\epsilon=0$ case with the help of formulas
(B.9), (B.8), and (B.5) from Appendix B.
The two integration regions in Eq\. (5.7b)
are entirely inside the region in Eq\. (5.7a), and shrink in proportion to
$\,a$.  Consequently, $L_1\to 0$ as $\,a\to 0$; hence the
integral of $L_1$ over $\,a\,$ and $\,G\,$ may be absorbed into the
finite terms in Eq\. (5.6).
$L_0$ may then be factored out of the integral, since it is
independent of $\,a\,$ and $\,G$.  The $\,a\,$ and $\,G\,$ integrations
are easily completed:  they produce the divergent factor
$\,\epsilon^{-1}$, plus a finite term, so that Eq\. (5.6) becomes
  \flushpar {\hsize=371pt \vskip -15pt $$
  \int_\Phi Y_1\ =\ \frac1{\dsize\epsilon} \int_0^1\! d\lambda
    \!\int_0^{\dsize\lambda}\! d\beta\ E_{{\dsize\epsilon}=0}
    \ \ + \ \text{finite terms}\,. \tag 5.8
  $$ \vskip 20pt }

The $Y_1$ terms arising from the other integrals in Tables 1 to 6
may be integrated in similar fashion.
After convergent terms have been discarded, each integral
factors into a trivial divergent integral,
times a finite integral which can be
evaluated at $\,\epsilon=0$ with the help of the formulas from Appendix B.
Adding the results of \it all \rm $Y_1$ integrations, we finally obtain the
total contribution to the \it divergent part \rm of $I$ from all $Y_1$ terms
(in Minkowski space):

$$ \align &\ \\ \vspace{-40pt}
  I_{Y_1}\ =\ \frac{\Gamma(4-2\omega)}{(4\pi)^{2\omega}}
    \,\Biggl[\,24\, n\!\!\!/ \left( \frac{m^2- p^2}{ n\!\cdot\! p}\right)
    \left[ L_2 \left( \frac{ p^2  }{ p^2-m^2} \right) -
           L_2 \left( \frac{ p^2+B}{ p^2-m^2} \right) \right] & \ +
  \\ \vspace{15pt}
    \bold N \cdot (\ 8,\ \ 22,-39,\ \ 89,\ \ \,0\ ) & \ +
  \\ \vspace{15pt}
    \left[ \frac{m^2}{ p^2} + \left( \dfrac{m^4}{ p^4}-1 \right)
    \ln \left( 1 - \frac{ p^2}{m^2} \right) - \ln(m^2) \right]
    \ \bold N \cdot (\ 0,\ \,-8,\ \ \ \,4,\ \,-4,\ \ \,0\ ) & \ +
  \\ \vspace{12pt}
    \left[ \left( \frac{m^2}{ p^2} - 1 \right)
    \ln \left( 1 - \frac{ p^2}{m^2} \right) - \ln(m^2) \right]
    \ \bold N \cdot (\ 8,\ \ \ \ 0,\ \ \ \,0,\ \ \ \,0,\ 24\ ) & \ +
  \\ \vspace{12pt}
    \left[ \left( \frac{m^2- p^2}{B+ p^2}\right)
    \ln \left( \frac{B+m^2}{m^2- p^2} \right) + \ln(B+m^2) \right]
    \ \bold N \cdot (\ 0,-16,\ \ 16,-32,\ 24\ ) &\ \Biggr], \endalign
$$ \flushpar {\hsize=371pt \vskip -20pt $$ \tag 5.9 $$ }
  \flushpar {\hsize=371pt \vskip -55pt $$ \align
  & \text{where}\qquad
    B\,\equiv\,-\,\frac{ n\!\cdot\! p\, n^*\!\!\cdot\! p}{n_0^2}\,,\qquad
    L_2(z) \,\equiv \int_0^1 \frac{\ln x \,dx}{x-1/z}\ ;
  \\ \vspace{15pt}
  & \text{the vector $\bold N$ is given by:} \qquad \bold N\ \equiv\ \biggl(
    m,\ p\!\!\!/,\ \frac{ n\!\cdot\! p\, n\!\!\!/^*}{ n\!\cdot\! n^*},
               \ \frac{ n^*\!\!\cdot\! p\, n\!\!\!/}{ n\!\cdot\! n^*},
    \ \frac{ p^2\, n\!\!\!/}{ n\!\cdot\! p} \biggr).\qquad \tag 5.10 \endalign
$$ }
\vskip 15pt
\leftline {\bf \ \ (b) Integrations over all parameter-space  }
\vskip 15pt
Finally, we must integrate the $Y_0$ terms over the finite parameters
$\,\lambda,\,\beta$, $G$, $b$, $h$, and $\,a$ (cf\. Eqs\. (5.1) and (4.2)).
Excluding the $\Gamma(2\epsilon)$ factor produced by the $A$ integration,
the integral of a $Y_0$ term will generally include a \it divergent \rm term
proportional to $\,\epsilon^{-1}$, plus a \it finite \rm term which is
independent of $\,\epsilon$, plus terms which go to zero as $\,\epsilon\to 0$.
Due to the $\Gamma(2\epsilon)$ factor, the \it divergent \rm term will
contribute to $I_{-2}$ (cf\. Eq\. (4.8)), while the \it finite \rm term will
contribute to $I_{-1}$.  Since we need both $I_{-2}$ and $I_{-1}$, we must
evaluate both the divergent and finite parts of the integral of $Y_0$.  The
divergent part comes from integration of $Y_0$ at its subdivergences, while
the finite part comes from integration over the whole region $\Phi$.
Since the subdivergences are part of the integral over the whole
region, we can extract both the divergent part \it and \rm the
finite part by integrating $Y_0$ over all of $\Phi$.

\medpagebreak
The functional form of the $Y_0$ terms is simpler than that of the $Y_1$
terms, because the former involve only the first term of the series in
Eq\. (4.9) and are, therefore, independent of $H$.  However, when the
denominator factor ${\,}_\wedge \,$ is present in the original integrand
$\,f(q,k)\ $ (i.e., when $\,G\neq 0$), integration
of the $Y_0$ terms is still complicated by the presence of the factors
$D_\parallel=a(1-a)-G^2$ and $D_\perp=\lambda h-G^2$ (see, for instance, Eq\.
(5.2a)).  In the integration of the $Y_1$ terms, we simplified these factors by
dropping all terms except those of lowest degree in $\{a,G\}$ or $\{1-a,G\}$.
This procedure is, unfortunately, of no use here, because higher-order terms
contribute to $I_{-1}$ via the \it finite \rm part of the integral of $Y_0$.
Our new strategy is, therefore, as follows:
we first simplify the $D_\parallel$ and $D_\perp$ factors by changing the
integration variables from $\,\{G,\,a,\,\lambda,\,\beta,\,b\}\,$ to
$\,\{V,\,U,\,X,\,\tau,\,t\}$, according to the following definitions:
  \flushpar {\hsize=371pt \vskip -12pt $$
  G = Va\,,\quad a=\frac1{U+V^2+1}\,,\quad \lambda=a(X+V^2)\,,\quad
  \beta=\tau a\,,\quad b=ta\,. \tag 5.11
  $$ \vskip 8pt } \flushpar

\medpagebreak
Note that this change of variables does not involve the parameter
$\,h$, and is, therefore, inapplicable to any integral containing
the factor $\,n\!\cdot\! k\,$ in the denominator of the original integrand
$\,f(q,k)$.  This restriction is not as severe as it might seem, because
we may always interchange $\,q\,$ and $\,k\,$ before integrating,
as mentioned in Section 3(c).  Hence the transformation (5.11)
may be applied to all integrals except those with \it both \rm
$\,n\!\cdot\! k\,$ and $\,n\!\cdot\! q\,$ in the denominator of $\,f$.
Fortunately, all such integrals of interest here (see Table 4) are harmless
anyway, since they give rise to $Y_0$ terms which are proportional to
$\,n^*\!\!\cdot\! n^*$, and hence vanish in the light-cone gauge.

\bigpagebreak
Under the change of variables (5.11), the boundaries $\,a=0\,$ and $\,a=1\,$
(at which all subdivergences occur) are transformed to $\,U\to \infty\,$ and
$\,U=V=0$, respectively, while the integrations in Eq\. (4.1) are transformed
according to:
  \flushpar {\hsize=371pt \vskip -6pt $$ \matrix
  \dsize\int_0^{\frac12}\!\! dG & \dsize\int_G^{1-G}\!\! da & \dsize
  \int_G^{1-\dsize a}\!\! d\lambda & \dsize\int_G^{\dsize\lambda}\!\! d\beta
  & \dsize\int_G^{\dsize a}\!\! db\,, \\ \vspace{6pt}
  \dsize\bold\downarrow & \dsize\bold\downarrow & \dsize\bold\downarrow &
  \dsize\bold\downarrow & \dsize\bold\downarrow \\ \vspace{6pt}
  \dsize\int_0^1\!\! dV & \dsize\int_{V-V^2}^\infty\!\! dU \ a^d
  & \dsize\int_{V-V^2}^U \!\! dX & \dsize\int_V^{X+V^2} \!\! d\tau
  & \dsize\int_V^1 \!\! dt\,, \endmatrix \tag 5.12
  $$ \vskip 12pt } \flushpar
where $\,d\,$ is the number of factors in the denominator of the original
integrand $\,f$, $\,a^d$ is the Jacobian determinant of the transformation,
and parameter integrations corresponding to absent
denominator factors are to be omitted, as already explained in Section 4(a).

\medpagebreak
Returning to our example, we now integrate Eq\. (5.2a) over the new
parameters.  From Eqs\. (5.11) and (3.7), we deduce $D_\parallel=a^2U$ and
$D_\perp=a^2X$, and since $\,\int db\,$ does not appear in Eq\. (4.2),
$\,\int dt\,$ is absent from the transformed integral:
  \flushpar {\hsize=371pt \vskip -12pt $$ \align \qquad
  \int_\Phi Y_0 \ &= \int_\Phi \frac{ n\!\!\!/^*\,a}
    {2D_\parallel^2 D_\perp^{1-\dsize\epsilon}}\ =
    \int_0^1\!\! dV \int_{V-V^2}^\infty\!\! dU \ a^5
    \int_{V-V^2}^U \!\! dX \int_V^{X+V^2} \!\! d\tau
    \ \frac { n\!\!\!/^*\,a^{-5+2\dsize\epsilon}}
                    {2\,U^2 X^{1-\dsize\epsilon}}\,,
  \\ \vspace{12pt}
  &= \,\frac{ n\!\!\!/^*}2\! \int_0^1\!\! dV\! \int_{V-V^2}^\infty\!\! dU
     \ a^{2\dsize\epsilon} \left[ \frac{U^{{\dsize\epsilon}-1}}{1+\epsilon}
      + \frac{(V\!-V^2)^{1+\dsize\epsilon}}{(1+\epsilon)\,\epsilon\,U^2}
      - \frac{V\!-V^2}{\epsilon\,U^{2-\dsize\epsilon}} \,\right].  \tag 5.13
  \endalign
  $$ \vskip 8pt } \flushpar
Since the old parameter $\,a\,$ is equal to the complicated
function $(U+V^2+1)^{-1}$, the new expressions $\,a^2U\,$ and $\,a^2X\,$ for
$D_\parallel$ and $D_\perp$ may not seem to represent a significant
improvement over the old ones.
However, when all factors of $\,a\,$ in the transformed integrand, including
the Jacobian $\,a^d$, are collected, it turns out that the net power of
$\,a\,$ is equal to $\,2\epsilon\,$ in every case.  This fact suggests that
we may be able to obtain the divergent and finite parts of the $Y_0$ integrals
by writing $\,(U+V^2+1)^{-2\dsize\epsilon}\,$ in series form, and keeping
only the first one or two terms.

\medpagebreak
In order to avoid the catastrophe of Eq\. (2.11), we need series which
converge uniformly over the whole domain of the $U,V$ integrations.
For $U>2$, the binomial series \vskip -25pt $$
  (U+ V^2+1)^{-2\dsize\epsilon} \,=\ U^{-2\dsize\epsilon} \left[ \,1\,-\,
  \frac{2\epsilon}1 \frac{(V^2+1)}U \,+\, \frac{2\epsilon}1
  \frac{(2\epsilon+1)}2 \frac{(V^2+1)^2}{U^2} \,-\ \dots\,\right]
$$ \flushpar {\hsize=371pt \vskip -25pt $$ \tag 5.14a
  $$ \vskip 0pt } \flushpar
is suitable, since its convergence accelerates with increasing $U$.
On the other hand for $U\leq 2$, an exponential series is appropriate:
\vskip -15pt $$
  (U +V^2+1 )^{-2\dsize\epsilon} \, = \ 1 \ - \ 2\epsilon\,\ln (U+V^2+1) \ +
   \ 2\epsilon^2\,(\ln (U+V^2+1))^2 \ - \ \dots \ .
$$ \flushpar {\hsize=371pt \vskip -30pt $$ \tag 5.14b
  $$ \vskip 0pt } \flushpar
In practice, one may avoid splitting up the integration region by using
Eq\. (5.14a) for integrals which diverge as $U\to\infty$ (e.g.: the first
term in brackets in Eq\. (5.13)), and Eq\. (5.14b)
for integrals which either diverge for finite $U$, or which do not diverge
at all (e.g.: the other two terms).
This strategy is valid, because in those parts of the integration region
where an integral converges, the finite part of the integral may always be
found by setting $\,\epsilon=0$, either before or after integration.  If we
set $\,\epsilon=0\,$ in Eq\. (5.14a) or (5.14b), all terms except for the
first one go to zero, and the two series become identical.

\medpagebreak
It turns out that, even at the points in the region of integration where a
$Y_0$ integral diverges, only the \it first term \rm of the appropriate
series contributes to the divergent and finite parts of the integral.  To see
why, we recall from Section 4(b) that the degree of the integrand of every
divergent finite-parameter integral is zero at the point in parameter-space
where the divergence occurs; in other words, every integral diverges like
$\,\int dx/x\,$ at $\,x=0$; if the degree were greater even by one,
the integral would not diverge.  In the series (5.14a), all terms but the
first are of higher degree in $\,1/U\,$ than the factor
$\,(U+V^2+1)^{-2\dsize\epsilon}$, so that
the integrals involving these terms will remain finite as $\,U\to \infty$.
The accompanying factors of $\,\epsilon\,$ in Eq\. (5.14a)
ensure that these terms do not even contribute to the \it finite \rm
part of the $Y_0$ integral as $\,\epsilon\to 0$.  Hence we may drop all
terms of the series, except the first one.  A similar argument applies
to the series (5.14b), where $\,\ln (U+V^2+1)\to 0$ linearly as $U,V\to 0$.

\medpagebreak
The preceding analysis tells us that
the complicated factor $\,(U+V^2+1)^{-2\dsize\epsilon}$,
appearing in the transformed integrands of the $Y_0$ integrals, may
always be replaced, either by $U^{-2\dsize\epsilon}$ or by 1.
Once this replacement has been made, integration of the $Y_0$ terms over
$\,t,\,\tau,\,X,$ and $U$ is trivial, while final integration over $V$
is easily accomplished with the help of the formula
  \flushpar {\hsize=371pt \vskip -8pt $$
  \int_0^1 V^C(1-V)^D\,dV\ =\ \frac{\Gamma(C+1)\,\Gamma(D+1)}{\Gamma(C+D+2)}\,;
  \tag 5.15
  $$ \vskip 8pt } \flushpar
here $C$ and $D$ are linear functions of $\,\epsilon$.

\medpagebreak
The results of the
integration of all $Y_0$ terms are shown in Appendix C.  When these results
are multiplied by the appropriate coefficients and then added, the total
contribution to the divergent part of
$I$ from all $Y_0$ terms reads as follows (in Minkowski space):
  \flushpar {\hsize=371pt \vskip -12pt $$
  I_{Y_0}\ =\ \frac{\Gamma(4-2\omega)}{(4\pi)^{2\omega}}
    \ \bold N\cdot \left(\frac{(\ 4,\ 4,-6,\ 14,\ 0\ )}{2-\omega}
    \ +\ (\ 8,-4,\ 11,\ 3,\ 0\ ) \right), \tag 5.16
  $$ \vskip 8pt } \flushpar
where $\,\bold N\,$ has already been defined in Eq\. (5.10).
To obtain the grand total of all divergent parts of $I$,
we simply add $I_{Y_0}$ from Eq\. (5.16) to $I_{Y_1}$ from Eq\. (5.9).
Thus (cf\. Eq\. (2.7))
  \flushpar {\hsize=371pt \vskip -12pt $$ \align
  I_0 \ &= \ ig^4\,T^b_{DA}T^a_{CD}T^b_{BC}T^a_{AB}\,I \\ \vspace{6pt}
        &= \ ig^4\,T^b_{DA}T^a_{CD}T^b_{BC}T^a_{AB}\,[I_{Y_0}+I_{Y_1}\ +
    \text{finite terms}\,]. \tag 5.17 \endalign
  $$ \vskip 8pt }

\newpage

\leftline{\bf 6. Concluding Remarks }
\vskip 15pt
In this paper we have developed a powerful new integration technique for
multi-loop integrals, called the \it matrix method\rm, and used it to compute
the divergent part of the overlapping two-loop fermion self-energy function
$\,i\Sigma\,$ in the light-cone gauge.  The overlapping self-energy diagram
(Fig\. 1) is the most challenging of the two-loop diagrams, especially in
a noncovariant gauge such as the light-cone gauge.  (The contributions from
Figs\. 2(a) to 2(d) are calculated in the sequel to this paper.)  For
completeness, we remind the reader of the one-loop result (Fig\. 3) in
the light-cone gauge [15]:
  \flushpar {\hsize=371pt \vskip -12pt $$
  i\Sigma_{1\text{-loop}} \ = \ \frac{g^2}{12\pi^2(2-\omega)}
    \left( p\!\!\!/+2m+2\,\frac{\, n^*\!\!\cdot\! p\, n\!\!\!/ - n\!\cdot\! p
    \, n\!\!\!/^*}{ n\!\cdot\! n^*}\right)\ +\text{finite terms.}\!\!\!\!
  \tag 6.1
  $$ \vskip 4pt } \flushpar

\medpagebreak
The main results for the overlapping self-energy function (cf\. Fig\. 1)
may be summarized thus:

\smallpagebreak
\item{(i)} The total divergent contribution, given by the sum of Eqs\.
(5.9) and (5.16), contains both simple and double poles.  The double-pole
term from Eq\. (5.16) reads
  \flushpar {\hsize=371pt \vskip -15pt $$
  \frac{\Gamma(4-2\omega)}{(4\pi)^{2\omega}(2-\omega)} \left(
    4m + 4p\!\!\!/ - 6\,\frac{ n\!\cdot\! p\, n\!\!\!/^*}{ n\!\cdot\! n^*} +
    14\,\frac{ n^*\!\!\cdot\! p\, n\!\!\!/}{ n\!\cdot\! n^*} \right), \tag 6.2
  $$ \vskip 8pt } \flushpar \item {}
which is seen to be \it local\rm, even off mass-shell.  The contributions to
$\,[i\Sigma]_{\dsize p^2\!=\!m^2}\! $ from results (6.1) and (6.2) are,
therefore, strictly local.  The coefficient of the single pole, on the other
hand, has also non-local terms, as seen from Eq\. (5.10).  (Both types of
poles are local on mass-shell.)  The locality of the double-pole term (6.2)
strongly suggests that the complete fermion-mass counterterm will likewise be
local, but confirmation will depend on the results from Figures 2(a) and 2(b).

\medpagebreak
\item{(ii)} The overlapping fermion self-energy integral contains a maximum
of \it seven \rm propagators.  Detailed analysis of the singularity structure
of the corresponding seven-parameter integrals in the ``user-friendly'' set
$S=\{A;\,\lambda,\beta,G,b,\mathbreak h,a\}$, Eq\. (3.4), reveals that the
\it first \rm simple pole originates from integration over the \it infinite
\rm parameter $A$, while the \it second \rm simple pole (and double pole
overall!) arises from integration over some of the \it finite \rm parameters
$\,\lambda,\,\beta$, $G$, $b$, $h$, and $\,a\,$ (subdivergences).

\medpagebreak
\item{(iii)} The matrix method is amazingly powerful, being applicable not
only to massive and massless integrals, but also to covariant integrals, as
depicted in Table 3, and to noncovariant integrals, i.e., those containing
the factors $\,(k\!\cdot\!n)^{-1}\,$ and/or $\,(q\!\cdot\!n)^{-1}\,$ (Tables
1, 2, 4, 5, and 6).  The success of the technique derives, quite simply,
from combining the $2\omega$-dimensional momentum vectors $\,q_\mu\,$ and
$\,k_\mu$, and then integrating over $4\omega$-dimensional Euclidean space
in a \it single \rm operation.  It is this compact procedure which yields
exact formulas (at an intermediate stage), whose analytic structure
simplifies the ensuing parameter integrations tremendously.  We note in
passing that the matrix method works equally well for axial-type gauges,
notably the temporal gauge ($n^2>0$) and the pure axial gauge ($n^2<0$).

\smallpagebreak
\item{(iv)} Although it has not been possible to corroborate our final
result against other calculations (there just aren't any!), we have
nevertheless been able to check all the covariant integrals listed in
Table 11.  And these agree exactly with the divergent parts of the double
covariant integrals currently used in computing radiative corrections in
the Standard Model [10,11].  Accordingly we feel reasonably confident that
our final answer for the overlapping integral $I$ in Eq\. (2.7) is correct.

\newpage
\centerline {\bf Acknowledgements }
\vskip 15pt
It gives us great pleasure to thank Sergio Fanchiotti for numerous helpful
discussions and for referring us to references such as refs\. [9],
[11] and [12].  We are also most grateful to him for using his computer
programs to check that the divergent parts of the \it covariant-gauge \rm
integrals in Table 11 agree with those employed in Standard Model
calculations.  One of us (G.L.) would like to thank Professors John Ellis
and Gabriele Veneziano, as well as the staff of the Theoretical Physics
Division, for their hospitality during his stay
at CERN, where part of the calculations were carried out.  Finally,
we should like to acknowledge support from NSERC of Canada under Grant
No\. A8063, and to thank the astute Referee for discovering a typographical
error in Eq\. (6.2).

\vskip 40pt
\centerline {\bf References}
\vskip 15pt
\item{[1] }G\. Leibbrandt,   Rev\. Mod\. Phys\. {\bf 59}, No.4 (1987) 1067;
 \newline A\. Bassetto, G\. Nardelli, and R\. Soldati, {\bf Yang-Mills
 Theories in Algebraic Non-covariant Gauges}, (World Scientific, Singapore,
 1991).
\item{[2] }P\. J\. Doust,   Ann\. Phys\. (N.Y.) {\bf 177} (1987) 169;
 P\. J\. Doust and J\. C\. Taylor,   Phys\. Lett\. {\bf B 197} (1987) 232.
\item{[3] }D\. M\. Capper, D\. R\. T\. Jones, and A\. T\. Suzuki,   Z\. Phys\.
 {\bf C 29} (1985) 585.
\item{[4] }A\. Smith,   Nucl\. Phys\. {\bf B 261} (1985) 285; Ph\. D\. thesis
 (University of California, Berkeley, 1985);   Nucl\. Phys\. {\bf B 267}
 (1986) 277.
\item{[5] }G\. Leibbrandt and S.-L\. Nyeo,   J\. Math\. Phys\. {\bf 27} (1986)
 627; G\. Leibbrandt,   Nucl\. Phys\. {\bf B 337} (1990) 87.
\item{[6] }A\. Andrasi and J\. C\. Taylor,   Nucl\. Phys\. {\bf B 375} (1992)
 341; Nucl\. Phys\. {\bf B 414} (1994) 856E.
\item{[7] }A\. Bassetto, I\. A\. Korchemskaya, G\. P\. Korchemsky, and
 G\. Nardelli,   Nucl\. Phys\. {\bf B 408} (1993) 62.
\item{[8] }G\. Leibbrandt, {\bf Noncovariant Gauges}, (World Scientific,
 Singapore, 1994).
\item{[9] }F\. A\. Berends and J\. B\. Tausk,   Nucl\. Phys\. {\bf B 421}
(1994)
 456; R\. Scharf and J\. B\. Tausk,   Nucl\. Phys\. {\bf B 412} (1994) 523;
 A\. I\. Davydychev and J\. B\. Tausk,   Nucl\. Phys\. {\bf B 397} (1993) 123;
 A\. I\. Davydychev, V\. A\. Smirnov and J\. B\. Tausk,   Nucl\. Phys\. {\bf B
 410} (1993) 325; D\. J\. Broadhurst,   Z\. Phys\. {\bf C 47} (1990) 115;
 D\. J\. Broadhurst, J\. Fleischer and O\. V\. Tarasov,   Z\. Phys\. {\bf C 60}
 (1993) 287; G\. Weiglein, R\. Mertig, R\. Scharf and M\. Bohm, in
 {\bf New Computing Techniques in Physics Research II}, ed\. D\. Perret-Gallix
 (World Scientific, Singapore, 1992) p\. 617;
 G\. Weiglein, R\. Scharf and M\. Bohm,   Nucl\. Phys\. {\bf B 416}
 (1994) 606.
\item{[10] }G\. Degrassi, S\. Fanchiotti and P\. Gambino, ``Two-loop
 next-to-leading $\,m_t\,$ corrections to the $\rho$ parameter'',
 preprint CERN-TH.7180/94, February 1994.
\item{[11] }S\. Bauberger, F\. A\. Berends, M\. Bohm, M\. Buza and
 G\. Weiglein, ``Calculation of two-looop self-energy in the electroweak
 Standard Model'', preprint INLO-PUB-11/94, June 1994;
 J\. Papavassiliou,   Phys\. Rev\. {\bf D 41} (1990) 3179; G\. Degrassi and
 A\. Sirlin,   Phys\. Rev\. {\bf D 46} (1992) 3104.
\item{[12] }A\. I\. Davydychev,   J\. Math\. Phys\. {\bf 32} (1991) 358;
 E\. E\. Boos and A\. I\. Davydychev,   Teor\. Mat\. Fiz\. {\bf 89} (1991) 56
 [Theor\. Math\. Phys\. USSR {\bf 89} (1991) 1052]; D\. Kreimer,   Phys\.
Lett\.
 {\bf B 273} (1991) 277; K\. G\. Chetyrkin, V\. A\. Ilyin, V\. A\. Smirnov and
 A\. Yu\. Taranov,   Phys\. Lett\. {\bf B 225} (1989) 411;
 J\. van der Bij and M\. Veltman,   Nucl\. Phys\. {\bf B 231} (1984) 205;
 F\. A\. Berends, G\. Burgers and W\. L\. van Neerven,   Nucl\. Phys\.
 {\bf B 297} (1988) 429; {\bf B 304} (1988) 921 (E);
 D\. Yu\. Bardin et al., \it in \rm Z physics at LEP 1, CERN 89-08 (1989) p\.
 89; W\. Beenakker, F\. A\. Berends and S\. C\. van der Marck,   Nucl\. Phys\.
 {\bf B 349} (1991) 323; {\bf B 367} (1991) 287;
 J\. van der Bij and F\. Hoogeveen,   Nucl\. Phys\. {\bf B 283} (1987) 477;
 R\. Barbieri, M\. Beccaria, P\. Ciafaloni, G\. Curci and
 A\. Vicere,   Phys\. Lett\. {\bf B 288} (1992) 95;   Nucl\. Phys\. {\bf B 409}
 (1993) 105.
\item{[13] }J\. L\. Rosner,   Ann\. Phys\. NY {\bf 44} (1967) 11;
 G\. 't Hooft and M\. Veltman, \it Diagrammar\rm, CERN preprint 73-9, 1973;
 M\. J\. Levine and R\. Roskies,   Phys\. Rev\. {\bf D 9} (1974) 421;
 E\. Mendels,   Nuovo Cimento {\bf A 45} (1978) 87; V\. K\. Cung, A\. Devoto,
 T\. Fulton and W\. W\. Repko,   Phys\. Rev\. {\bf D 18} (1978) 3893;
 A\. A\. Vladimirov,   Teor\. Mat\. Fiz\. {\bf 43} (1980) 210
 [Theor\. Math\. Phys\. USSR {\bf 43} (1980) 417];
 A\. E\. Terrano,   Phys\. Lett\. {\bf B 93} (1980) 424;
 F\. V\. Tkachov,   Phys\. Lett\.
 {\bf B 100} (1981) 65; D\. R\. T\. Jones and J\. P\. Leveille,
   Nucl\. Phys\. {\bf B 206} (1982) 473; D\. I\. Kazakov,   Phys\. Lett\.
 {\bf B 133} (1983) 406; N\. Marcus and A\. Sagnotti,   Nucl\. Phys\.
 {\bf B 256} (1985) 77; G\. Leibbrandt, Rev\. Mod\. Phys\. {\bf 47} (1975)
 849 (see Section 7.A).
\item{[14] }R\. Tarrach, Nucl\. Phys\. {\bf B 183} (1981) 384.
\item{[15] }G\. Leibbrandt and S.-L\. Nyeo, Phys\. Lett\. {\bf B 140} (1984)
 417.
\item{[16] }S\. Mandelstam, Nucl\. Phys\. {\bf B 213} (1983) 149.
\item{[17] }G\. Leibbrandt, Phys\. Rev\. {\bf D 29} (1984) 1699.
\item{[18] }S\. Weinberg, Phys\. Rev\. {\bf 118} (1960) 838.

\vskip 40pt
\leftline {\bf Figure Captions }
\vskip 15pt
\item{Fig\. 1\ } Two-loop overlapping quark self-energy function (solid
lines denote quarks, wavy lines gluons).
\item{Fig\. 2\ } Other two-loop diagrams for the quark self-energy.
\item{Fig\. 3\ } One-loop quark self-energy function computed in ref\. [15].

\newpage
\leftline {\bf Appendix A: \ \ Useful momentum-space integrals. }
\vskip 15pt
The following formulas were derived from Eqs\. (3.5) to (3.14).
Note that $\,\bold z\,\equiv\,(k_4,\,q_4,\,k_3,\,q_3,\,\dots\,)^\top$.
\flushpar {\hsize=371pt \vskip -10pt $$ \align
J[k_\mu] \ &= \ r_\mu\,J[1],
\qquad\qquad\qquad\qquad\qquad\qquad\qquad \tag A.1 \\ \vspace{15pt}
J[q_\mu] \ &= \ s_\mu\,J[1],
\qquad\qquad\qquad\qquad\qquad\qquad\qquad \tag A.2 \\ \vspace{15pt}
J[ n^*\!\cdot\! q\,k_\mu] \ &= \ \left(
 n^*\!\cdot\! s \,r_\mu-\dfrac{Gn^*_\mu}{2AD_\parallel}\right)J[1],
\ \ \qquad\qquad\qquad\qquad \tag A.3 \\ \vspace{15pt}
J[ n^*\!\cdot\! q\,q_\mu] \ &= \ \left(
 n^*\!\cdot\! s\,s_\mu+\dfrac{an^*_\mu}{2AD_\parallel}\right)J[1],
\ \ \qquad\qquad\qquad\qquad \tag A.4 \\ \vspace{15pt}
J[ n^*\!\cdot\! q\, n\!\cdot\! k\, k\!\!\!/] \ &=
\ \left( n^*\!\cdot\! s\, n\!\cdot\! r\, r\!\!\!/\,
+\,\dfrac{\beta\, n^*\!\cdot\! p\, n\!\!\!/\,+\,2Gn_0^2
( r\!\!\!/+ r\!\!\!/_\parallel)}{2AD_\parallel}\right)J[1], \tag A.5
\\ \vspace{15pt}
J[ n^*\!\cdot\! q\, n\!\cdot\! k\, q\!\!\!/] \ &=
\ \left( n^*\!\cdot\! s\, n\!\cdot\! r\, s\!\!\!/\,
+\,\dfrac{b\, n\!\cdot\! p\, n\!\!\!/^*\,+\,2Gn_0^2
( s\!\!\!/+ s\!\!\!/_\parallel)}{2AD_\parallel}\right)J[1],\qquad \tag A.6
\\ \vspace{15pt}
J[ n^*\!\cdot\! q\, q\!\cdot\! k] \ &=
\ \left( n^*\!\cdot\! s\, s\!\cdot\! r\,
+\,\dfrac{b\, n^*\!\cdot\! p\,+\,O(G)}{2AD_\parallel}\right)J[1],
\qquad\qquad \tag A.7
\\ \vspace{25pt}
\text{where}\qquad  J[1] \ &= \ \left(\frac{\pi}A\right)^{4-2\dsize\epsilon}
\frac{\bold e^{-AH}}{D_\parallel D_\perp^{1-\dsize\epsilon}}\,,
\qquad (\epsilon\,\equiv\,2-\omega),  \tag A.8
\\ \vspace{15pt}
H \,\equiv\, \frac{C-\bold B\!\cdot\!\bold m}A \ &=
\ (b+\beta -G)( p^2+m^2)-(b r+\beta s)\!\cdot\! p\,.  \tag A.9
\endalign $$ } \flushpar
Since $J$ is a linear functional, we can use $\,J[q_\mu]\,$ to deduce
integrals such as $\,J[ n^*\!\cdot\! q]\,$ and $\,J[ q\!\!\!/]$, etc.

\newpage
\leftline {\bf Appendix B: \ \ Formulas for $Y_1$ integration}
\vskip 15pt
Integration of $Y_1$ terms requires integration over finite parameters
of expressions involving $\,H_0\equiv H_{{\dsize a}\to 0}\,$ and
$\,H_1\equiv H_{{\dsize a,h}\to 1}$.  $H_0$ is given by Eqs\. (5.5),
which in turn were derived from Eq\. (A.9).  We may similarly derive
$\,H_1\,=\,p_\parallel^2\,b\,(R-b-bS)\,$, where $R$ and $S$ are
the same as in Eqs\. (5.5).

\bigpagebreak
Formulas (B.1) to (B.11) may be obtained by elementary means.  Since
$\, p^2 = p_\parallel^2 + p_\perp^2 $, it follows from Eqs\. (5.5) that
$\ p_\parallel^2\,(R-S-1)=m^2$.

{\hsize=371pt \vskip -10pt $$ \align
\int_0^1\!\! d\lambda\,\Bigl[\ln H_0\Bigr]_{\dsize\beta\!=\!\lambda}\ \ &=
\ \ \ln(m^2)-2+(R\!-\!S)\ln\left[\frac{R-S}{R\!-\!S\!-\!1}\right], \tag B.1
\\ \vspace{15pt}
\int_0^1\!\! db\ \ln H_1 \quad &= \ \ \ln(m^2)-2+\frac R{S\!+\!1}
\ln\left[\frac R{R\!-\!S\!-\!1}\right], \tag B.2
\\ \vspace {15pt}
2\int_0^1\!\! db\ b\,\ln H_1 \ \ &= \ \ \ln(m^2)-1+
 \frac{R^2}{(S\!+\!1)^2}\ln\left[\frac R{R\!-\!S\!-\!1}\right]
 -\frac R{S\!+\!1},\quad \tag B.3
\\ \vspace {15pt}
\int_0^1\!\! d\lambda \!\int_0^{\dsize\lambda}\! d\beta
 \ \frac{\beta\,p_\parallel^2}{\lambda H_0} \ \ &=
 \ \ L_2\left(\frac{S+1}R\right)-L_2\left(\frac SR\right), \tag B.4
\\ \vspace {10pt}
& \text{where} \qquad L_2(z)\ \equiv\,\int_0^1 \frac{\ln x\ dx}{x-1/z}\,.
\endalign
$$ \vskip 12pt } \flushpar
Formula (B.4) was derived with the help of the substitution
$\,\beta = y\lambda\,$, followed by integration over $\,\lambda\,$ and then
integration by parts over $\,y$.  The same substitution is helpful in the
derivation of some of the integrals (B.5) to (B.11) in Table 9.

\medpagebreak
In formulas (B.5) to (B.11), $\ \bold W \ \equiv $ \vskip -15pt
$$\left(\ln(m^2),\ 1,\ \ln\left[\frac R{R\!-\!S\!-1}\right],
\ (R\!-\!S)\ln\left[\frac{R-S}{R\!-\!S\!-1}\right],
\ L_2\left[\frac{S\!+\!1}R\right]-L_2\left[\frac{S}R\right] \right),
$$ \flushpar
{\hsize=371pt \vskip -20pt $$ \tag B.12 $$ } \flushpar
so that, for example, the right-hand side of formula (B.5) is \vskip -15pt
$$
\ln(m^2)\ -\ 4\ +\ (R\!-\!S)\ln\left[\frac{R-S}{R\!-\!S\!-1}\right]\ +
\ R\,L_2\left[\frac{S\!+\!1}R\right]\,-\,R\,L_2\left[\frac{S}R\right].
$$
\vskip 25pt
\centerline{\underbar{Table 9}} \vskip 40pt
\flushpar --------------------------------------------
\!\!\!\! ---------------------------------------------------
\vskip -69pt
$$ \matrix E & \Biggl| &&
 \dsize\int_0^1\! d\lambda \int_0^{\dsize\lambda}\! d\beta \ \ E &&&&&&&
\\ & \Bigl| &&&&&&&&& \\
 \dfrac{\ln H_0}{\lambda} \!\!\!& \Biggl| &\!\!\! \bold W\!\cdot\!\biggl(
 1, \!\!\!\!& -4 \!\!\!\!&,&\!\!\!\! 0 \!\!\!\!&,&\!\!\!\! 1
 \!\!\!\!&,&\!\!\!\! R \ \ \biggr) & (\text{B}.5)
\\  & \Bigl| &&&&&&&&& \\ \vspace{-3pt}
 2\,\dfrac{\beta\,\ln H_0}{\lambda^2} \!\!\!& \Biggl| &\!\!\! \bold W\!\cdot\!
 \biggl( 1, \!\!\!\!& -3 \!\!\!\!&,&\!\!\!\! \dfrac{R^2}{S(S\!+\!1)}
\!\!\!\!&,&\!\!\!\! 1-\dfrac RS \!\!\!\!&,&\!\!\!\! 0\ \ \biggr) & (\text{B}.6)
\\  & \Bigl| &&&&&&&&& \\ \vspace{-3pt}
 4\,\dfrac{\beta\,\ln H_0}{\lambda} \!\!\!& \Biggl| &\!\!\! \bold W\!\cdot\!
 \biggl( 1, \!\!\!\!& S-3R-2 \!\!\!\!&,&\!\!\!\! \dfrac{-2\,R^2}{S\!+\!1}
 \!\!\!\!&,&\!\!\!\! 3R\!-\!S \!\!\!\!&,&\!\!\!\! 2R^2 \biggr) & (\text{B}.7)
\\  & \Bigl| &&&&&&&&& \\ \vspace{-3pt}
 \dfrac{\beta^2\,p_\parallel^2}{\lambda^2 H_0} \!\!\!& \Biggl| &\!\!\!
 \bold W\!\cdot\!\biggl(0, \!\!\!\!& 0 \!\!\!\!&,&\!\!\!\! \dfrac R{S(S\!+\!1)}
 \!\!\!\!&,&\!\!\!\! -\dfrac1S \!\!\!\!&,&\!\!\!\! 0\ \ \biggr) & (\text{B}.8)
\\  & \Bigl| &&&&&&&&& \\ \vspace{-3pt}
 \dfrac{\beta^2\,p_\parallel^2}{\lambda H_0} \!\!\!& \Biggl| &\!\!\!
 \bold W\!\cdot\!\biggl(0, \!\!\!\!& -1 \!\!\!\!&,&\!\!\!\! \dfrac{-R}{S\!+\!1}
 \!\!\!\!&,&\!\!\!\! 1 \!\!\!\!&,&\!\!\!\! R\ \ \biggr) & (\text{B}.9)
\\  & \Bigl| &&&&&&&&& \\ \vspace{-3pt}
 2\,\dfrac{\beta^3\,p_\parallel^2}{\lambda^2 H_0} \!\!\!& \Biggl| &\!\!\!
 \bold W\!\cdot\! \biggl(0, \!\!\!\!& \dfrac R{S\!+\!1}-1 \!\!\!\!&,&\!\!\!\!
 \dfrac{R^2}{S(S\!+\!1)^2} \!\!\!\!&,&\!\!\!\! 1-\dfrac RS \!\!\!\!&,&\!\!\!\!
 0\ \ \biggr) & (\text{B}.10)
\\  & \Bigl| &&&&&&&&& \\ \vspace{-3pt}
 2\,\dfrac{\beta^3\,p_\parallel^2}{\lambda H_0} \!\!\!& \Biggl| &\!\!\!
 \bold W\!\cdot\! \biggl(0, \!\!\!\!& \dfrac R{S\!+\!1}+\!S\!-\!3R\!-\dfrac12
 \!\!\!\!&,&\!\!\!\! \dfrac{-R^2(2S\!+\!3)}{(S\!+\!1)^2} \!\!\!\!&,&\!\!\!\!
 3R\!-\!S \!\!\!\!&,&\!\!\!\! 2R^2 \biggr) & (\text{B}.11)
\\ & \bigl| &&&&&&&&& \\
\endmatrix $$
\vskip -12pt \flushpar --------------------------------------------------
\!\!\!\! --------------------------------------------- \vskip 10pt

\newpage
\leftline {\bf Appendix C: \ \ Results of $Y_0$ integration }
\vskip 15pt
Each entry in the following tables, when multiplied by
$\,\pi^{2\omega}\Gamma(4-2\omega)\,$ times the Euclidean-vector expression
at the right-hand side of its row of the table, gives the $Y_0$ portion of
the divergent part of the integral of the expression at the left-hand side
divided by the denominator at the bottom.
Denominators are represented in the notation
of definitions (3.15), and $\ \epsilon\,\equiv\,2-\omega\,$.

\baselineskip=19pt
\settabs 9 \columns

\vskip 12pt
\centerline{\underbar{Table 10}}
\vskip -8pt \+&------------------&------------------&------------------
&------------------&------------------&------------------&------------------\cr
\vskip -18pt \+&------------------&------------------&------------------
&------------------&------------------&------------------&------------------\cr
\vskip -14pt \+&& $\|$ & $|$ & $|$ & $|$ & $\|$ & $\|$ \cr
\vskip -14pt \+& \ \ \ \ \ $k_\mu$ & \ \ \ \ \ \ \ --1 & \ \ \ \ \ \ \ --1 &
\,\ --$\,\tfrac1{2\dsize\epsilon}+\tfrac12$ &
\,\ --$\,\tfrac1{2\dsize\epsilon}+\tfrac12$ &
\ \ \ \ \ \ \ 0 & \ \ \ \ \ $n_\mu^*/ n\!\cdot\! n^*$ \cr
\vskip -14pt \+&& $\|$ & $|$ & $|$ & $|$ & $\|$ & $\|$ \cr
\vskip -14pt \+&------------------&------------------&------------------
&------------------&------------------&------------------&------------------\cr
\vskip -14pt \+&& $\|$ & $|$ & $|$ & $|$ & $\|$ & $\|$ \cr
\vskip -14pt \+& \ \ \ \ \ $q_\mu$ &
\ \ \ $\,\tfrac1{\dsize\epsilon}+1$ &\ \ \ $\,\tfrac1{\dsize\epsilon}+1$ &
\ \ \ $\,\tfrac1{\dsize\epsilon}-1$ &\ \ \ $\,\tfrac1{\dsize\epsilon}-1$ &
\ \ \ $\,\tfrac2{\dsize\epsilon}-2$ &
\ \ \ \ \ $n_\mu^*/ n\!\cdot\! n^*$ \cr
\vskip -14pt \+&& $\|$ & $|$ & $|$ & $|$ & $\|$ & $\|$ \cr
\vskip -14pt \+&------------------&------------------&------------------
&------------------&------------------&------------------&------------------\cr
\vskip -14pt \+&& $\|$ & $|$ & $|$ & $|$ & $\|$ & $\|$ \cr
\vskip -14pt \+& \ $ n\!\!\!/\, k\!\cdot\! q$ &&&&&
\ \ \ $\,\tfrac1{\dsize\epsilon}-1$ &
\ \ $ n\!\!\!/\, n^*\!\!\cdot\! p/ n\!\cdot\! n^*$ \cr
\vskip -14pt \+&& $\|$ & $|$ & $|$ & $|$ & $\|$ & $\|$ \cr
\vskip -14pt \+&------------------&------------------&------------------
&------------------&------------------&------------------&------------------\cr
\vskip -18pt \+&------------------&------------------&------------------
&------------------&------------------&------------------&------------------\cr
\vskip -14pt \+&& $\|$ & $|$ & $|$ & $|$ & $\|$ & $\|$ \cr
\vskip -14pt \+&&\ \ \ $\tfrac1{2\dsize\epsilon}+1$ &
\ \ \ $\tfrac1{2\dsize\epsilon}+1$ &\ \ \ $\,\tfrac1{\dsize\epsilon}-1$ &
\ \ \ $\tfrac1{2\dsize\epsilon}-\tfrac12$ &\ \ \ $\,\tfrac1{\dsize\epsilon}-1$
 &\ \ $ n\!\!\!/^*\, n\!\cdot\! p/ n\!\cdot\! n^*$ \cr
\vskip -14pt \+&& $\|$ & $|$ & $|$ & $|$ & $\|$ & $\|$ \cr
\vskip -14pt \+&&------------------&------------------
&------------------&------------------&------------------&------------------\cr
\vskip -14pt \+&& $\|$ & $|$ & $|$ & $|$ & $\|$ & $\|$ \cr
\vskip -14pt \+& \ $ n\!\cdot\! k\, q\!\!\!/$ &
\ \ \ \ \ \ \ \ 2 &\ \ \ \ \ \ \ \ 6 &
\ \ \ $\,\tfrac2{\dsize\epsilon}-1$ &
\ \ \ $\,\tfrac2{\dsize\epsilon}+1$ &
\ \ \ \ \ \ \ 0 & \ \ \ \ \ --$\, p\!\!\!/_\parallel/8$ \cr
\vskip -14pt \+&& $\|$ & $|$ & $|$ & $|$ & $\|$ & $\|$ \cr
\vskip -14pt \+&&------------------&------------------
&------------------&------------------&------------------&------------------\cr
\vskip -14pt \+&& $\|$ & $|$ & $|$ & $|$ & $\|$ & $\|$ \cr
\vskip -14pt \+&& \ \ \ \ \ \ \ \ 2 &\ \ \ \ \ \ \ \ 6 &
\ \ \ $\,\tfrac2{\dsize\epsilon}-3$ &
\ \ \ $\,\tfrac2{\dsize\epsilon}-1$ &
\ \ \ \ \ \ \ 0 & \ \ \ \ \ --$\, p\!\!\!/_\perp/8$  \cr
\vskip -14pt \+&& $\|$ & $|$ & $|$ & $|$ & $\|$ & $\|$ \cr
\vskip -14pt \+&------------------&------------------&------------------
&------------------&------------------&------------------&------------------\cr
\vskip -18pt \+&------------------&------------------&------------------
&------------------&------------------&------------------&------------------\cr
\vskip -14pt \+&& $\|$ & $|$ & $|$ & $|$ & $\|$ & $\|$ \cr
\vskip -14pt \+&&\ \ \ \ \ \ \ \ 1 &
\ \ \ $\tfrac1{2\dsize\epsilon}+\tfrac12$ &&&&
\ \ $ n\!\!\!/\, n^*\!\!\cdot\! p/ n\!\cdot\! n^*$ \cr
\vskip -14pt \+&& $\|$ & $|$ & $|$ & $|$ & $\|$ & $\|$ \cr
\vskip -14pt \+&&------------------&------------------
&------------------&------------------&------------------&------------------\cr
\vskip -14pt \+&& $\|$ & $|$ & $|$ & $|$ & $\|$ & $\|$ \cr
\vskip -14pt \+& \ $ n\!\cdot\! k\, k\!\!\!/$ &
\ \ \ \ \ \ \ 11 &\ \ \ \ \ \ \ \ 9 &&&&
\ \ \ \ \ --$\, p\!\!\!/_\parallel/8$ \cr
\vskip -14pt \+&& $\|$ & $|$ & $|$ & $|$ & $\|$ & $\|$ \cr
\vskip -14pt \+&&------------------&------------------
&------------------&------------------&------------------&------------------\cr
\vskip -14pt \+&& $\|$ & $|$ & $|$ & $|$ & $\|$ & $\|$ \cr
\vskip -14pt \+&&\ \ \ \ \ \ \ \ 5 &\ \ \ \ \ \ \ \ 3 &&&&
\ \ \ \ \ --$\, p\!\!\!/_\perp/8$ \cr
\vskip -14pt \+&& $\|$ & $|$ & $|$ & $|$ & $\|$ & $\|$ \cr
\vskip -14pt \+&------------------&------------------&------------------
&------------------&------------------&------------------&------------------\cr
\vskip -18pt \+&------------------&------------------&------------------
&------------------&------------------&------------------&------------------\cr
\vskip -14pt \+&& $\|$ & $|$ & $|$ & $|$ & $\|$ & $\|$ \cr
\vskip -14pt \+&& $\|$ nq ${}_\wedge$Kk & $|$ n Q${}_\wedge$Kk & $|$
 nqQ${}_\wedge$K & $|$ nqQ${}_\wedge$ k & $\|$\,nqQ Kk & $\|$ \cr

\vskip 15pt
\centerline{\underbar{Table 11}}
\vskip -8pt\+&------------------&------------------&------------------
&------------------&------------------&------------------&------------------\cr
\vskip -18pt \+&------------------&------------------&------------------
&------------------&------------------&------------------&------------------\cr
\vskip -14pt \+&& $\|$ & $|$ & $|$ & $|$ & $\|$ & $\|$ \cr
\vskip -14pt \+& \ \ \ \ \ \ 1 &
\ \ \ $\,\tfrac1{\dsize\epsilon}+1$ &\ \ \ $\,\tfrac1{\dsize\epsilon}+1$ &
\ \ \ $\,\tfrac1{\dsize\epsilon}+1$ &\ \ \ $\,\tfrac1{\dsize\epsilon}+1$ &
\ \ \ $\ \tfrac2{\dsize\epsilon}$ & \ \ \ \ \ \ \ 1 \cr
\vskip -14pt \+&& $\|$ & $|$ & $|$ & $|$ & $\|$ & $\|$ \cr
\vskip -14pt \+&------------------&------------------&------------------
&------------------&------------------&------------------&------------------\cr
\vskip -14pt \+&& $\|$ & $|$ & $|$ & $|$ & $\|$ & $\|$ \cr
\vskip -14pt \+& \ \ \ \ \ \ $ q\!\!\!/$ &
\ \ \ $\,\tfrac2{\dsize\epsilon}+1$ &\ \ \ $\,\tfrac6{\dsize\epsilon}+7$ &
\ \ \ $\,\tfrac4{\dsize\epsilon}+2$ &\ \ \ $\,\tfrac4{\dsize\epsilon}+6$ &
\ \ \ $\ \tfrac8{\dsize\epsilon}$ &\ \ \ \ \ \ $\, p\!\!\!//8$ \cr
\vskip -14pt \+&& $\|$ & $|$ & $|$ & $|$ & $\|$ & $\|$ \cr
\vskip -14pt \+&------------------&------------------&------------------
&------------------&------------------&------------------&------------------\cr
\vskip -18pt \+&------------------&------------------&------------------
&------------------&------------------&------------------&------------------\cr
\vskip -14pt \+&& $\|$ & $|$ & $|$ & $|$ & $\|$ & $\|$ \cr
\vskip -14pt \+&& $\|$ \ q ${}_\wedge$Kk & $|$ \  Q${}_\wedge$Kk & $|$
 \ qQ${}_\wedge$K & $|$ \ qQ${}_\wedge$ k & $\|$\,\ qQ Kk & $\|$ \cr

\vskip 15pt
\centerline{\underbar{Table 12}}
\vskip -8pt\+&&------------------&------------------
&------------------&------------------&------------------&------------------\cr
\vskip -18pt \+&&------------------&------------------
&------------------&------------------&------------------&------------------\cr
\vskip -14pt \+&&& $\|$ & $|$ & $|$ & $\|$ & $\|$ \cr
\vskip -14pt \+&& \ \ \ \ \ \ $ n\!\!\!/$ &
\ \ \ $1/\epsilon$ & \ \ \ \ \ \ 0 &
\ \ \ $2/\epsilon$ &\ \ $(4/\epsilon)$\,--\,4 &
\ \ $ n\!\!\!/\, n^*\!\!\cdot\! p/ n\!\cdot\! n^*$ \cr
\vskip -14pt \+&&& $\|$ & $|$ & $|$ & $\|$ & $\|$ \cr
\vskip -14pt \+&&------------------&------------------
&------------------&------------------&------------------&------------------\cr
\vskip -18pt \+&&------------------&------------------
&------------------&------------------&------------------&------------------\cr
\vskip -14pt \+&&& $\|$ & $|$ & $|$ & $\|$ & $\|$ \cr
\vskip -14pt \+&&& $\|$ n \ ${}_\wedge$Kk & $|$ nq ${}_\wedge$K & $|$
 n Q${}_\wedge$ k & $\|$ n Q Kk & $\|$ \cr

\baselineskip=20pt
\newpage
\leftline {\bf Appendix D: \ \ Computational summary of a typical
overlapping integral. }
\vskip 15pt
We summarize the main steps in the computation of the double integral,
corresponding to the fifth row and second-last column of Table 1,
Section 3(c):
  \flushpar {\hsize=371pt \vskip -10pt $$
  I_E[f]\ \equiv\ \int_E d^{2\omega}q \int_E d^{2\omega}k \ f\,, \qquad
  \text{with}\quad f=\,\frac{q\!\!\!/}{\text{nqQ${}_\wedge$ k}}\,. \tag D.1
  $$ \vskip 5pt } \flushpar
Using definitions (3.15), we find that
  \flushpar {\hsize=371pt \vskip -18pt $$
  I_E[f]\ =\ \int_E \int_E  \,\frac {q\!\!\!/\ \ d^{2\omega}q\ d^{2\omega}k}
  {\,n\!\cdot\!q \ q^2\,[(p-q)^2+m^2]\ [(p-k-q)^2+m^2]\ k^2}\,. \tag D.2
  $$ \vskip 8pt }

\item{1.} \,First, we perform the exponential parametrization (2.9) with
Schwinger parameters $\,\alpha_1,\,\dots\,,\alpha_7\ \,(\alpha_5=\alpha_7=0)$;
we then replace these parameters by the
``user-friendly'' set $\{A;\lambda,\beta,G,b,h,a\}$, defined in Eqs\. (3.4).
Thus (cf\. Eq\. (4.2)),
  \flushpar {\hsize=371pt \vskip -18pt $$ \align
  I_E[f]\ &= \int_0^\infty\!\! d\alpha_1\ \dots \int_0^\infty\!\! d\alpha_4
    \ \int_0^\infty\!\! d\alpha_6
    \  J\left[\frac{-n^*\!\!\cdot\!q\ q\!\!\!/} {n_0^2}\right]
             _{\dsize\alpha_5\to 0\atop\dsize\alpha_7\to 0}, \tag D.3
  \\ \vspace{10pt}
  &= \int_\Phi \int_0^\infty\!\! A^4\,dA
     \ J\left[\frac{-n^*\!\!\cdot\! q\ q\!\!\!/} {n_0^2}\right]
     _{\dsize b\to G\atop\dsize h\to a}\,, \tag D.4
  \\ \vspace{10pt} \qquad\quad
  \text{with } &J \text{ defined by Eq\. (3.3), and }  \int_\Phi
    \equiv \int_0^\frac12\! dG \int_G^{1-G}\! da \int_G^{1-{\dsize a}}\!
    d\lambda \int_G^{\dsize\lambda}\! d\beta\,. \endalign
  $$ \vskip 5pt }

\item{2.} \,Next, we perform the $4\omega$-dimensional \it momentum
integration \rm in $J$ with the help of formulas (A.4) and (A.8).  In this
way we obtain (cf\. Eq\. (4.4)):
  \flushpar {\hsize=371pt \vskip -15pt $$
  I_E[f]\ =\ -\,\frac{\pi^{4-2\dsize\epsilon}}{n_0^2} \int_\Phi
    \int_0^\infty\!\! dA\,\left[ \frac{A^{2\dsize\epsilon}\ \bold e^{-AH}}
    {D_\parallel D_\perp^{1-\dsize\epsilon}} \left( n^*\!\!\cdot\! s\ s\!\!\!/
    \,+\,\frac{a n\!\!\!/^*}{2AD_\parallel}\right) \right], \tag D.5
  $$ \vskip 5pt} \item{}
with $D_\parallel,\,D_\perp,\,r,\,s\,$ and $H$ being defined in
Eqs\. (3.7), (3.14) and (A.9). \,(In Eq\. (D.5), and all subsequent
equations, $\,b\to G,\ h\to a,\,$ and $\,\epsilon\equiv 2-\omega\,$.)

\bigpagebreak
\item{3.} Integration over the infinite (type I) parameter $A$ yields the
\it first \rm simple pole (cf\. Eq\. (4.6)):
  \flushpar {\hsize=371pt \vskip -15pt $$
  I_E[f]\ =\ -\,\frac{\pi^{4-2\dsize\epsilon}}{n_0^2} \int_\Phi \bigl[
    \,\Gamma(2\epsilon)\ J_0\ +\ \Gamma(1+2\epsilon)\ J_1\,\bigr], \tag D.6
  $$ \vskip 8pt} \item{}
with $J_0$ and $J_1$ given in Eqs\. (4.7).  The term
with $\Gamma(2\epsilon)$ diverges as $\,\epsilon\to 0$ $(\omega\to 2)$.
Additional \it subdivergences \rm will arise from integration over $\Phi$.

\bigpagebreak
\item{4.} Both $J_0$ and $J_1$ include factors of $H^{-2\dsize\epsilon}$.
According to Section 4(b), there are no subdivergences at the zeros of $H$;
therefore, we use the exponential series for $H^{-2\dsize\epsilon}$
given in Eq\. (4.9) to get (cf\. Eq\. (5.1)):
  \flushpar {\hsize=371pt \vskip -12pt $$
  I_E[f]\ =\ -\,\frac{\pi^{4-2\dsize\epsilon}}{n_0^2} \int_\Phi
    \bigl[\,\Gamma(2\epsilon)\ Y_0\ +
    \ \Gamma(1+2\epsilon)\ Y_1\ \,+ O(\epsilon)\,\bigr]\,, \tag D.7
  $$ \vskip 8pt} \item{}
with $Y_0$ and $Y_1$ shown in Eqs\. (5.2).  As discussed in Section 5,
we require both the \it finite \rm and \it divergent \rm parts of
$\int_\Phi Y_0$, but only the divergent part of $\int_\Phi Y_1$.

\bigpagebreak
\item{5(a).} In this particular example, all subdivergences occur at $\,a=G=0$.
(In general, according to Section 4(b), subdivergences may also occur at
$\,a=1$, $G=0$.)  To find the divergent part of $\int_\Phi Y_1$, therefore, we
integrate only the portion of $Y_1$ of least degree in $\{a,G\}$ (cf\. Eqs\.
(5.3) to (5.7)):
  \flushpar {\hsize=371pt \vskip -12pt $$ \align
  \int_\Phi Y_1\ &= \int_0^{\frac12}\! dG \int_G^{1-G}\! da
     \int_G^{1-{\dsize a}}\!\!\! d\lambda \!\int_G^{\dsize\lambda}\!\!
     d\beta\ [\,a^{{\dsize\epsilon}-2}\,E\ +O(a^{{\dsize\epsilon}-1})\,]\,,
     \tag D.8
  \\ \vspace{12pt}
  &= \int_0^{\frac12}\! dG \int_G^{1-G} a^{{\dsize\epsilon}-2}\, da
     \ \int_0^1 d\lambda \int_0^{\dsize\lambda} d\beta\ E \quad +
     \ \text{finite terms}\,, \tag D.9 \endalign
  $$ \vskip 8pt } \item{}
with $E$ being defined in Eq\. (5.4).  Since $E$ is independent of $\,a\,$ and
$\,G$,
  \flushpar {\hsize=371pt \vskip -5pt $$
  \int_\Phi Y_1\ =\ \frac1{\dsize\epsilon} \int_0^1\! d\lambda
    \!\int_0^{\dsize\lambda}\! d\beta\ E_{{\dsize\epsilon}=0}
    \ \ + \ \text{finite terms}\,. \tag D.10
  $$ \vskip 5pt }

\item{5(b).} Integration over $\,\lambda\,$ and $\,\beta\,$ with the help of
Eqs\. (B.9), (B.8), and (B.5) yields:
  \flushpar {\hsize=371pt \vskip -15pt $$ \align
  \int_\Phi Y_1\ &=\ \frac{\bold W}{\dsize\epsilon} \cdot \Biggl[\,
    \frac{ n^*\!\!\cdot\! p\, p\!\!\!/_\parallel} { p_\parallel^2}
          \left(0,\, -1,\, \frac{-R}{S\!+\!1},\, 1,\, R \right)\,-\,
     \frac{ n\!\!\!/^*}2  (1,\, -4,\, 0,\, 1,\, R )
  \\ \vspace{10pt} & \qquad
     +\ \frac{ n^*\!\!\cdot\! p\, p\!\!\!/_\perp} { p_\parallel^2}
          \left(0,\, 0,\, \frac R{S(S\!+\!1)},\, \frac{-1}S,\, 0 \right)
    \,\Biggr]\ \ + \ \text{finite terms}\,, \tag D.11
  \endalign
  $$ \vskip 5pt } \item{}
with $\bold W$ being defined in Eq\. (B.12), and $R$ and $S$ in Eqs\. (5.5).

\bigpagebreak
\item{6(a).} To integrate $Y_0$ over $\Phi$, we first change the integration
variables from $\{a,\,G,$ $\lambda,\,\beta\}$ to $\{U,\,V,\,X,\,\tau\}$ via
Eqs\. (5.11).  From Eqs\. (3.7) we deduce $D_\parallel=a^2U$ and
$D_\perp=a^2X$, so that the integral of Eq\. (5.2a) over $\Phi$ becomes
(cf\. Eq\. (5.13)):
  \flushpar {\hsize=371pt \vskip -20pt $$ \align \qquad
  \int_\Phi Y_0 \ &= \int_0^1\!\! dV \int_{V-V^2}^\infty\!\! dU \ a^5
    \int_{V-V^2}^U \!\! dX \int_V^{X+V^2} \!\! d\tau
    \ \frac { n\!\!\!/^*\,a^{-5+2\dsize\epsilon}}
                    {2\,U^2 X^{1-\dsize\epsilon}}\,,
  \\ \vspace{12pt}
  = \,& \frac{ n\!\!\!/^*}2\! \int_0^1\!\! dV\! \int_{V-V^2}^\infty\!\! dU
    \ a^{2\dsize\epsilon} \left[ \frac{U^{{\dsize\epsilon}-1}}{1+\epsilon}
    + \frac{(V\!-V^2)^{1+\dsize\epsilon}}{(1+\epsilon)\,\epsilon\,U^2}
    - \frac{V\!-V^2}{\epsilon\,U^{2-\dsize\epsilon}} \,\right],\quad \tag D.12
  \endalign
  $$ \vskip 5pt } \item{}
with $\,a=(U+V^2+1)^{-1}$, according to Eqs\. (5.11).

\bigpagebreak
\item{6(b).} As explained in Section 5(b), we obtain the divergent and finite
parts of $\int_\Phi Y_0$ by replacing $\,a^{2\dsize\epsilon}\,$ with
$\,U^{-2\dsize\epsilon}\,$ in those terms whose integrals diverge as
$U\to\infty$, and with 1 in the other terms.  Thus,
  \flushpar {\hsize=371pt \vskip -15pt $$ \align \qquad \quad
  \int_\Phi Y_0 \ &=\,\frac{ n\!\!\!/^*}2\! \int_0^1\!\! dV \,
    \left[\,\frac{(V\!-V^2)^{-\dsize\epsilon}}{\epsilon(1+\epsilon)} \,+\,
            \frac{(V\!-V^2)^{\dsize\epsilon}}{\epsilon(1+\epsilon)} \,-\,
            \frac{(V\!-V^2)^{\dsize\epsilon}}{\epsilon(1-\epsilon)} \,\right]
    \ +\,O(\epsilon)\,,
  \\ \vspace{12pt}
  &=\,\frac{ n\!\!\!/^*}2\,\left(\frac1{\dsize\epsilon}-1\right)
    \ +\,O(\epsilon)\,. \tag D.13  \endalign
  $$ \vskip 5pt } \item{}
The divergent part of $I_E[f]$ may be obtained by combining
Eqs\. (D.7), (D.11), and (D.13).

}\bye